\documentclass {article}

\usepackage[T1]{fontenc}
\usepackage{amsmath}
\usepackage{amsfonts}
\usepackage{amssymb}
\usepackage{latexsym}
\usepackage[english]{babel}
\usepackage[dvips]{graphicx,color}
\usepackage{latexsym}
\usepackage{graphicx}
\usepackage{graphics}
\usepackage{color}
\usepackage{fancyhdr}
\usepackage{fancyhdr}
\usepackage{pstricks}

\newcommand{\be}{\begin{equation}}
\newcommand{\ee}{\end{equation}}
\newcommand{\beq}{\begin{equation}}
\newcommand{\eeq}{\end{equation}}
\newcommand{\bea}{\begin{eqnarray}}
\newcommand{\eea}{\end{eqnarray}}
\newcommand{\nn}{\nonumber}
\newcommand{\ba}{\begin{eqnarray}}
\newcommand{\ea}{\end{eqnarray}}
\def\bc{\begin{center}}
\def\ec{\end{center}}

\def\lab{\label}

\begin{document}

\begin{titlepage}

\vspace{2in}

\begin{centering}

{\Large {\bf Wilson loops stability in the
  gauge/string correspondence}}

\vspace{.3in}

Ra\'ul E. Arias  and Guillermo A. Silva \\
\vspace{.2 in}
{\it IFLP-CONICET and Departamento de F\'{\i}sica\\ Facultad de Ciencias Exactas, Universidad Nacional de La Plata\\
CC 67, 1900,  La Plata, Argentina}\\
\vspace{.4in}

{\bf Abstract} \\

\end{centering}

~

\noindent We study the stability of some classical string worldsheet
solutions employed for computing the potential energy between two
static fundamental quarks in confining and non-confining gravity
duals. We discuss  the fixing of the diffeomorphism
invariance of the string action, its relation with the fluctuation orientation and
the interpretation of the quark mass
substraction worldsheet needed for computing the potential energy in smooth (confining)
gravity background. We consider various dual gravity  backgrounds and show by a numerical
analysis the existence of instabilities under linear fluctuations
for classical string embedding solutions having positive length function derivative $L'(r_0)>0$.
Finally we make a brief discussion of 't Hooft loops in non-conformal backgrounds.

\end{titlepage}

\section{Introduction}

The proof of confinement in non-abelian gauge theories from first
principles remains to date unsolved. The strong-coupling aspect of
the phenomenon precludes from attacking it with standard QFT
perturbative techniques. Nevertheless a criteria for confinement was
proposed long ago by Wilson \cite{wilson}. The criteria states that
an area law behavior for the so called Wilson loop
indicates confinement of the chromoelectric flux tubes (QCD string).
In particular, an area law behavior for a rectangular (infinite
strip) spacetime contour corresponding to a static quark-antiquark
pair indicates a linear confining potential  between quarks.

In the latest years, the gauge/string correspondence \cite{malda}
has provided new insights into the confinement phenomenon. The
crucial observation is that the string aspect of the chromoelectric
flux tube manifests in the dual perspective by the appearance of a
holographic dimension \cite{malda},\cite{polyakov}. A string theory
prescription for computing Wilson loops was proposed in
\cite{maldawilson}: the gauge theory loop is to be though at
infinity in the radial holographic coordinate and the Wilson loop
for fundamental quarks is defined by an open string whose endpoints
lie on the loop at infinity. In the large $N_c$ limit ('t Hooft
limit) the QCD string self-interactions vanish \cite{thooft} and the
dual gravity prescription for computing the potential energy between
quarks in a given gauge theory amounts to finding a minimal surface
in the corresponding gravity dual.

The canonical computation for the potential energy between a pair of
fundamental static quarks (rec\-tan\-gular loop) involves a U-shaped string extending in
the holographic direction where the gauge theory quarks separation $L$
translates into the separation between the fixed open string
endpoints located at infinity. As the quarks separation varies, the
string worldsheet explores the holographic direction, therefore, the minimum
radial position $r_0$ reached by the string depend on
the endpoints
separation distance $L$.  This procedure was applied to a number of
paradigmatic examples and gave results consistent with gauge theory
expectations, in particular a theorem
stating sufficient conditions for confining backgrounds was
proved in \cite{sonnen} (see \cite{sonnwilson} for a review).
Extensions to higher gauge group representations and 't Hooft loops
were analyzed and proposed in \cite{groo},\cite{hk},\cite{gomis},
they involve higher dimensional $D$-branes with or without
worldvolume gauge fields turned on.

In many applications to dual gravity backgrounds the prescription
\cite{maldawilson} has been applied at the classical zeroth order
level to establish confinement, phase transitions or transport
properties \cite{sonnwilson},\cite{gubser}, only recently has the
stability of some classical string embeddings been studied
\cite{kmt},\cite{pufu},\cite{avramis} (see also
\cite{cg},\cite{sonnkin},\cite{forste},\cite{dgt}). One of the
motivations for the stability analysis, in generalized situations, has been the appearance of
multiple classical embedding solutions for given boundary conditions
\cite{pufu},\cite{avramis},\cite{argyres} (see also \cite{Brandh}) signaled
by the presence of extrema in the length function $L(r_0)$ (see \cite{cotrone} for related recent work). 
The presence of a maximum separation length
was interpreted as dual to the occurrence of screening.
The aim of the present work is to show that whenever one has a  $L'(r_0)>0$ branch of
solutions, they are unstable. We will confirm this statement by explicitly
showing the existence, in particular gravity backgrounds, of
unstable ($\omega^2<0$) modes for the $L'(r_0)>0$ branches.  It is worth mentioning that this is
a satisfying result since the expected physical behavior for the $L(r_0)$ relation from the gauge/string
correspondence is to have $L'(r_0)<0$. We will mention briefly an analysis of 't Hooft loops computations
in non-conformal gravity duals where instabilities also arise.

Along the way we will discuss various aspects of the classical embeddings:
the first one regards the
physical interpretation of the configuration employed for
obtaining a finite potential energy between quarks in smooth gravity backgrounds, the second regards the diffeomorphism
invariance of the string action and its relation to the possible
gauge choices for the orientation of the in-plane fluctuations, the third one is the relation between
instabilities of the string embedding and the $L(r_0)$ relation
between the separation of the string endpoints at infinity $L$ and
the maximum depth reached by the string probe in the holographic
direction $r_0$. It was proved in \cite{avramis} that the presence
of an extremum in the $L(r_0)$ relation leads to the existence of  a
zero mode for the longitudinal fluctuations, signaling an instability. We will
confirm this fact by explicitly computing the lowest fluctuation
modes in a number of gravity duals examples.

The paper is organized as follows: in section \ref{wil} we review
the prescription for computing Wilson loops from gravity
backgrounds. In section \ref{bkg} we describe the backgrounds we
will study and compute their length and energy functions. In section \ref{stab} we perform a quadratic
fluctuations analysis and compute numerically the lowest fluctuation modes. In section \ref{scho}
we reanalyze the result of  section \ref{stab} transforming the fluctuations
equations of motion into a Schrodinger problem. In section \ref{thooftloop} we briefly discuss the
't Hooft loop case and in
section \ref{concl} we summarize our conclusions. We conclude with two appendices
with technical details.

\section{Wilson loops and string solutions}
\label{wil}

\subsection{Static string U-shaped embeddings}
\label{wilson}
The starting point for Wilson loop computations,
in the large $N_c,\lambda$ ('t Hooft limit),
from gravity duals with metric $g_{\mu\nu}$ is the Nambu-Goto action\footnote{Generically
one should take into account contributions from $B_2$ background fields but
for the ansatz we will consider they do not contribute to (\ref{NG}).}
\be
S=\frac\eta{2\pi\alpha'}\int d\tau d\sigma\sqrt{\eta\, h}\,.
\label{NG}
\ee
Here $h=\det h_{\alpha\beta}$, $h_{\alpha\beta}=g_{\mu\nu}\partial_\alpha X^\mu
\partial_\beta X^\nu$ is the induced metric on the string worldsheet,
$\partial_\alpha=\partial/\partial\xi^\alpha$ with
$\xi^\alpha=\{\tau,\sigma\}$ the string worldsheet coordinates and
$X^\mu$ run over the target space coordinates. A sign $\eta$
accounts for possible Euclidean ($\eta=+$) and (timelike) Lorentzian
($\eta=-$) configurations. The class of metrics  we consider take the form
\be
ds^2=-g_{t}(r)dt^2+g_{x}(r)dx_i^2+g_{r}(r)dr^2+g_{ab}(r,\theta)d\theta^a d\theta^b\,.
\label{gravity}
\ee
The $t,x_i\,(i=1,2,3)$ coordinates represent the gauge theory
coordinates, $r$ is the bulk holographic coordinate and $\theta_a\,(a,b=1,..,5)$ are
additional angular coordinates parametrizing a compact 5d space $\Sigma_5$. The
potential energy between quarks involves solving the NG action for
(timelike) worldsheets corresponding to strings whose endpoints at
infinity lie on the loop to be computed, typically the endpoints are
kept separated by a constant distance $L$ in one of the $x_i$ coordinates
which we call $x$ (see \cite{sonnwilson} for a review). We
start analyzing static embeddings of the form $t(\tau),x(\sigma),r(\sigma)$, with all other coordinates
fixed to constants\footnote{Generalizations considering moving quarks on the boundary, relevant
for QGP applications, are straightforward and rephrased in terms of non-diagonal
terms in the metric\cite{pufu},\cite{argyres}.}. Placing the anzats into the action
leads to the correct equations of motion, one therefore has
\ba
S&=&-\frac1{2\pi\alpha'}\int d\tau d\sigma \sqrt{g_{t}(r)\,\dot t^2\left( g_{x}(r)\,\acute x^2+g_{r}(r)\,\acute r^2\right)}\nn\\
 &=&-\frac1{2\pi\alpha'}\int dt d\sigma \sqrt{g_{t}(r)\left( g_{x}(r)\,\acute x^2+g_{r}(r)\,\acute r^2\right)}\nn\\
 &=&-\frac{\cal T}{2\pi\alpha'}\int d\sigma \sqrt{ f^2(r)\,\acute x^2+g^2(r)\,\acute r^2}\,.
 \label{effec}
\ea
where $g^2(r)=g_{t}(r)g_{r}(r)$ and $f^2(r)=g_{t}(r)g_{x}(r)$.
The reparametrization invariance of (\ref{NG}) factorizes the temporal extension of the loop $\cal T$ and reduces the Wilson
loop computation to finding a geodesic in the effective 2-dimensional geometry
\be
ds^2_{eff}=f^2(r)dx^2+g^2(r)dr^2\,.
\label{geff}
\ee
The conserved charge associated to
$x$-translations in (\ref{effec}) is
\be
\frac{f^2(r)\, \acute x(\sigma)}{\sqrt{f^2(r)\acute x(\sigma)^2+g^2(r)\acute r(\sigma)^2}}=A
\ee
from which one obtains
\be
\acute x(\sigma)=\pm A{\frac{g(r)}{f(r)}}\frac1{\sqrt{f^2(r)-A^2}}\,\acute r(\sigma)\,.
\label{eqxr}
\ee
Reparametrization invariance guarantees that (\ref{eqxr}) solves the $r$-equation
of motion. Calling $r_0$ the point given by $f(r_0)=A$, (\ref{eqxr}) can
rewritten as \cite{sonnen}
\be
\frac{dx}{dr}=\pm {\frac{g(r)}{f(r)}}\frac{f(r_0)}{\sqrt{f^2(r)-f^2(r_0)}}\,.
\label{sig}
\ee
The boundary conditions at infinity for the string endpoints separation
are $\Delta x|_{r=\infty}=L$. From (\ref{sig}) one
notices that the string reaches the boundary in an orthogonal way. Two natural gauge choices that appear in the
literature are: $x(\sigma)=\sigma$
($x$-gauge) or $r(\sigma)=\sigma$ ($r$-gauge). The first choice ($x$-gauge) has the benefit of providing
a complete parametrization of the embedding $r(x)$ when imposing $x\in[-L/2,L/2]$ and $r(\pm L/2)=\infty$
(the tip of the string is conventionally chosen to be at
$(r,x)=(r_0,0)$). Making $x\to t$, equation (\ref{sig}) can be understood as a zero energy motion in
a potential $U(r)$ given by
\be
U(r)=\frac{f^2(r)(f^2(r)-f^2(r_0))}{g^2(r)f^2(r_0)},
\ee
the point
$r_0$ is therefore easily seen to be the minimum value in the holographic coordinate reached  by the string.
The second choice ($r$-gauge) gives a double valued
$x(r)$ relation when imposing $r\in[r_0,\infty)$ and $x(\infty)=\pm L/2$.
Nevertheless in several examples leads to closed analytical expressions for $x(r)$ and moreover, when
computing fluctuations around the static solution, drastically simplify the equations of motion since no fluctuations in the
metric components $g_{\mu\nu}$ should be taken into account (see \cite{sonnen},\cite{avramis}). One should keep in
mind that the tip of the string is a special point in the $r$-gauge since we must patch there the two branches
corresponding to the $\pm$ signs in (\ref{sig}) (see the following section).

Integrating (\ref{eqxr}) we arrive to the important $L(r_0)$ length function,
\be
L(r_0)=2\int_{r_0}^\infty
\frac{g(r)}{f(r)}\frac{f(r_0)}{\sqrt{f^2(r)-f^2({r_0})}}\,dr\,.
\label{generalL}
\ee
Assuming that $f(r),f'(r)>0$ the lower limit $r_0$ in (\ref{generalL}) is generically integrable\footnote{A
zero  $f'(r_0)=0$  leads to a $r_0$-dependent logarithmic
divergence in (\ref{generalL}) (see \cite{piai} for a generic discussion on divergences in the length function).}.
Note that a finite lhs in (\ref{generalL}) demands that $g/f^2$ should decay
at infinity faster than $1/r$. In the following sections we will be
interested in the relation (\ref{generalL}). The derivative $L'(r_0)$ can be computed as
follows \cite{avramis} (see also the recent work \cite{piai}),
\bea
\frac{L'(r_0)}2&=&-\left.\frac{g(r)}{\sqrt{f^2(r)-f^2({r_0})}}\right|_{r\to r_0}
+f'(r_0)\int_{r_0}^\infty  \frac{f(r)g(r)}{(f^2(r)-f^2(r_0))^{\frac32}}\,dr\nn\\
&=&-\left.\frac{g(r)}{\sqrt{f^2(r)-f^2({r_0})}}\right|_{r\to r_0}+f'(r_0)\int_{r_0}^\infty dr \frac{g(r)}{f'(r)} \frac d{dr}
\left(-\frac1{\sqrt{f^2(r)-f^2(r_0)}}\right)\nn\\
&=&-\left.\frac{f'(r_0)g(r)}{f'(r)\sqrt{f^2(r)-f^2({r_0})}}\right|_{r\to\infty}+\int_{r_0}^\infty dr \frac{f'(r_0)}{\sqrt{f^2(r)-f^2(r_0)}}\frac d{dr}
\left(\frac{g(r)}{f'(r)}\right)\,,\nn
\eea
where we have integrated by parts when passing from the second to the third line. Since the first term
in the rhs of the third line vanishes in all the backgrounds we consider, we obtain
\be
{L'(r_0)}=2\int_{r_0}^\infty dr \frac{f'(r_0)}{\sqrt{f^2(r)-f^2(r_0)}}\frac d{dr}
\left(\frac{g(r)}{f'(r)}\right)\,.\label{Lprima}
\ee

The energy of the $q\bar q$ configuration was  proposed in  \cite{maldawilson} to be
given by the length of the string
solution (\ref{eqxr}) in the effective 2d metric (\ref{geff}),
\be
E=\frac 1{2\pi\alpha'}\int d\sigma\sqrt{f^2(r)\acute
x(\sigma)^2+g^2(r)\acute r(\sigma)^2}
\label{energy}
\ee
The expressions for the energy in the $x$- and $r$-gauges, using (\ref{eqxr}) are
\ba
E(r_0)&=&\frac 1{2\pi\alpha'}\int_{-L/2}^{L/2} dx\,\frac{f^2(r(x))}{f(r_0)}\\
&=&\frac 1{\pi\alpha'}\int_{r_0}^\infty
dr\frac{g(r)f(r)}{\sqrt{f^2(r)-f^2(r_0)}}\,.
\label{er}
\ea
The energy computed
by expression (\ref{energy}) diverges due to the infinite extension
of the string\footnote{The divergence in (\ref{er}) is generically $r_0$-independent,
an additional $r_0$-dependent divergence might appear in (\ref{generalL}) and (\ref{energy})
whenever the string stretches to regions where $f'(r_0)=0$.}. The interpretation for this divergence is that
(\ref{energy}) contains, in addition to the potential energy between
quarks, the self energy (mass) of the external quarks
\cite{maldawilson}. In order to obtain a meaningful quantity and get
the potential between quarks we should compare (\ref{energy}) with
respect to a reference state taking care to substract a $r_0$
independent quantity. It is customary to take
the length of a straight string stretching from infinity all the way
down to the interior of the bulk spacetime along the $r$ coordinate,
with all other coordinates set to constants, as the ``bare'' quark mass. Calling $r=r_{min}$ the
minimum allowed value for the radial coordinate in the geometry (\ref{gravity}),
either because of presence of a horizon (e.g. $AdS$ in Poincare
coordinates or thermal BH backgrounds) or because the spacetime ends
in a regular fashion (e.g. Witten $AdS$ soliton, Maldacena N\'u\~nez
and Klebanov-Strassler backgrounds), the quark mass takes the form\footnote{See
the discussion at the end of sections \ref{adsglobal} and \ref{mnsol} regarding the interpretation of
the reference state in smooth gravity backgrounds.}
\be
m_q=\frac1{2\pi\alpha'}\int_{r_{min}}^\infty g(r)\, dr\, .
\label{se}
\ee
The potential energy between quarks obtained from
(\ref{energy}) after substracting the quarks self energy (\ref{se})
is
\bea
E_{q\bar q}(r_0)&=&E(r_0)-2\,m_q\nn\\
&=&\frac 1{\pi\alpha'}\left[\int_{r_0}^\infty \frac{g(r)f(r)}{\sqrt{f^2(r)-f^2(r_0)}}\,dr-\int_{r_{min}}^\infty g(r)\, dr\right]\,.
\label{enerren}
\eea
Eliminating $r_0$ from (\ref{generalL}) and (\ref{enerren}) we obtain the gauge/string proposal for the potential energy
between quarks  in the planar large 't Hooft limit $V_{\sf string}(L)$.
In the following sections we will plot this relation in several examples and analyze its functional form. For completeness we
compute the derivative of (\ref{enerren}), one has
\be
E'_{q\bar q}(r_0)=\frac1{\pi\alpha'}\left[-\left.\frac{g(r)f(r)}{\sqrt{f^2(r)-f^2({r_0})}}\right|_{r=r_0}+\int_{r_0}^\infty dr \frac{f(r)g(r)f(r_0)f'(r_0)}{(f^2(r)-f^2(r_0))^{\frac32}}\right]\,.\nn
\ee
Using the first line of (\ref{Lprima}) one obtains \cite{avramis}
\be
E'_{q\bar q}(r_0)=\frac1{2\pi\alpha'}f(r_0)\,L'(r_0)\Rightarrow\frac{dE_{q\bar q}}{dL}=\frac1{2\pi\alpha'}f(r_0)\,,
\label{eprima}
\ee
where $r_0$ in the last expression  should be understood as the function $r_0(L)$ obtained by inverting (\ref{generalL}).

\vspace{1.5mm}

We end this section quoting some conditions that must be satisfied by any potential pretending to describe the
interaction between physical quarks. The so called `concavity' conditions proved in \cite{bachas} are
\be
\frac{dV}{dL}>0,~~~~~~\frac{d^2V}{dL^2}\leq0\,.
\label{convexity}
\ee
This conditions hold independently of the gauge group and the details of the matter sector. The physical interpretation
of (\ref{convexity}) is that  the force between the quark-antiquark pair is: (i) always attractive and (ii) a non increasing function of their
separation distance. From (\ref{generalL}),(\ref{enerren}),(\ref{eprima})
we find that the string proposal $V_{\sf string}(L)$ gives \cite{Brandh}
\be
\frac{dV_{\sf string}}{dL}=\frac{dE_{q\bar q}}{dr_0}\frac{dr_0}{dL}=\frac{1}{2\pi\alpha'} f(r_0)
,~~~~~~~~\frac{d^2V_{\sf string}}{dL^2}=\frac{1}{2\pi\alpha'}\left(\frac{dL}{dr_0}\right)^{-1}{f'(r_0)}\,.
\label{concavity}
\ee
The first condition is always met in dual gravity backgrounds since by definition $f(r)>0$.
Although in all our examples $f'(r)>0$,
the second condition might fail whenever $L'(r_0)$ is positive. We will present cases where this non-physical behavior
appears and show that precisely in those circumstances the string embedding solution (\ref{sig})-(\ref{generalL})
is unstable under small perturbations. This last statement was the motivation of the present work.

\subsection{Stability analysis of classical string embeddings}
\label{stability}
We will study in this section the stability of the classical solution $(r_{\sf cl}(\sigma),x_{\sf cl}(\sigma))$
given by (\ref{sig})-(\ref{generalL}) under small (linear) perturbations. A general fluctuation around the embedding
solution can be written as
\be
X^\mu=(\tau,x_{\sf cl}(\sigma)+\delta x_1(\tau,\sigma),\delta x_2(\tau,\sigma),\delta x_3(\tau,\sigma),
r_{\sf cl}(\sigma)+\delta r(\tau,\sigma),\theta^a+\delta\theta^a(\tau,\sigma))\, .\label{embedding}
\ee
We can  use the diffeomorphism invariance of the action to fix $t=\tau$ and forget about the $t$-equation of motion.
For the class of metrics considered in (\ref{gravity}), the $\delta x_2$ and $\delta x_3$ fluctuations decouple
and as expected satisfy the same equation of motion, the
$\delta\theta^a$ fluctuations mix among themselves for generic compact manifolds
and lead to five eom  (we will not analyze the angular fluctuations in the present work and we consistently set them to zero),
finally the $\delta x_1$ and $\delta r$ fluctuations result mixed
in two coupled equations.

\vspace{3.5cm}

\begin{figure}[h]
\begin{minipage}{7cm}
\hspace{2.5cm}
\pscircle[linewidth=.3pt](2.5,0){.3}
\psline[linewidth=.3pt]{->}(0,-3)(0,3)
\psline[linewidth=.3pt]{<-}(0,0)(3.5,0)
\psarc[linewidth=.5pt,linestyle=dashed](0,0){2.5}{270}{90}
\pscurve(0,2.5)(0.05,2.505)(0.3,2.9)
(0.7,1.8)(1.2,2.7)(1.6,1.2)(2.1,2.3)(2.48,.8)
(2.5,0)(2.48,-.8)(2.1,-2.3)(1.6,-1.2)
(1.2,-2.7)(0.7,-1.8)(0.3,-2.9)(0.05,-2.505)(0,-2.5)
\rput[bI](-0.3,2.9){$x$}
\rput[bI](3.7,-0.3){$r$}
\rput[bI](3,-0.4){$r_0$}
\rput[bI](2.9,0.3){$?$}
\rput[bI](-.2,2.2){$\frac L2$}
\rput[bI](-.4,-2.7){$-\frac L2$}
\vspace{3cm}
\caption{{\sf $r$-gauge fixing}: The dashed line represents the classical embedding
over which we perturb. At the tip $r_0$, the fluctuation is oriented along the string,
and therefore not physical.}
\label{r-gauge}
\end{minipage}
\   \  \
\hfill
\begin{minipage}{7cm}
\hspace{2cm}
\psline[linewidth=.3pt]{->}(0,-3)(0,3)
\psline{->}(2.5,0)(3.2,0)
\psline[linewidth=.3pt]{<-}(0,0)(4,0)
\psarc[linewidth=.5pt,linestyle=dashed](0,0){2.5}{270}{90}
\pscurve(0,2.5)(0.5,2.5)(1.3,2.3)(0.5,2.1)(2.4,1.7)(1,1.3)(2.9,0.8)
(1.5,0.5)(3.1,0.2)(3.2,0)(3.1,-0.2)(1.5,-0.5)(2.9,-0.8)(1,-1.3)(2.4,-1.7)(0.5,-2.1)(1.3,-2.3)
(0.5,-2.5)(0,-2.5)
\rput[bI](-0.3,2.9){$x$}
\rput[bI](3.4, 0.2){$\delta r_0$}
\rput[bI](3.9,-0.3){$r$}
\rput[bI](-.2,2.2){$\frac L2$}
\rput[bI](-.4,-2.7){$-\frac L2$}
\vspace{3cm}
\caption{{\sf $x$-gauge fixing}: The dashed line represents the classical embedding we are perturbing.
This is a physical gauge choice all over the embedding solution, the plotted  (even) fluctuation
changes the position of the tip.}
\label{xgauge}
\end{minipage}
\end{figure}

It is easily shown that the $r$- and $x_1$-equations of motion are
proportional. The remaining diffeomorphism
should therefore be used to fix the orientation of the $(\delta r,\delta x_1)$ vector (in-plane fluctuation) at each point of
the solution $(r_{\sf cl}(\sigma),x_{\sf cl}(\sigma))$. After imposing a gauge constraint
one equation describes the fluctuations in the $(r,x_1)$-plane and we end up with a well posed system of differential equations.
The physical gauge choice ($n$-gauge)
would be to orient the fluctuation along the normal direction to the classical embedding $(r_{\sf cl}(\sigma),x_{\sf cl}(\sigma))$
and generically this means to account for fluctuations on both $x_1$ and $r$ coordinates. Two other natural possibilities considered
in the literature correspond to fixing $\delta x_1(\tau,\sigma)=0$ ($x$-gauge) or $\delta r(\tau,\sigma)=0$ ($r$-gauge). The $n$-
and $x$-gauge fixings, as mentioned in the previous section, parametrize the fluctuations along the whole classical
embedding but the equations of motion are lengthier because the $\delta r$ fluctuations result in additional contributions to the eom
arising from changes in the metric \cite{sonnkin},\cite{avramis}. Note that at first sight the $n$-
and $x$-gauge fixings appear to allow for the oscillation of the tip, while the $r$-gauge fixing should not (more on this in the following).
We will choose to work in the $r$-gauge, this means fixing
$\delta r(\tau,\sigma)=0$ and work with simpler equations defined on half of the embedding. It will then be mandatory to analyze
the boundary conditions to be imposed at the tip $(r_0,0)$ of the embedding in order to get a meaningful solution. Moreover, since
we are considering fluctuations along the $x_1$ coordinate, precisely at the tip the $\delta x_1$ fluctuation  is
oriented along the string worldvolume and not transverse to it:
\emph{ the $\delta x_1$ fluctuation in the $r$-gauge  is therefore not physical at the tip
of the embedding}.
The required additional analysis developed in \cite{pufu},\cite{avramis} will be discussed below. At last,
another advantage of the $r$-gauge is that it gives closed expressions for the linearized fluctuations equations of motion
(see (\ref{SL})-(\ref{spp})), the
isometry along $x_1$ implies that upon computing fluctuations over the classical solution (\ref{sig})-(\ref{generalL}), we do not
need the explicit analytic solution $x_{\sf cl}(r)$, what contribute to the equations
of motion is its derivative (\ref{sig}) (as an example compare (\ref{xg}) with (\ref{cg})).
We would like to quote the work \cite{pufu} where $\delta r$ fluctuations
in the $\delta x_1=0$ gauge fixing were
considered over (half) the classical embedding parametrized in the $r$-gauge (see below eqn. (\ref{x1}),(\ref{rr})).

In the following we will be concerned with the equations of motion for the $\delta x_i$
fluctuations, they can be obtained by placing the
ansatz
\be
t=\tau,~~~~x_1=x_{\sf cl}(r)+\delta x_1(t,r),~~~~x_2=\delta
x_2(t,r),~~~~x_3=\delta x_3(t,r),~~~~r=\sigma \, ,
\label{Avramis}
\ee
into the action (\ref{NG}). Expanding to second order in the fluctuations one
obtains
\bea
2\pi\alpha'{\cal
L}^{(2)}&=&\frac1{g(r)f(r)\sqrt{f^2(r)-f^2(r_0)}}
\left[\phantom{\sum}\!\!\!\!\!\!\!\!h^2(r)\,(f^2(r)-f^2(r_0))\,(\delta \dot x_1)^2-(f^2(r)-f^2(r_0))^2(\delta \acute x_1)^2\right.\nn\\
&&\left.+f^2(r)h^2(r)((\delta \dot x_2)^2+(\delta \dot x_3)^2)-f^2(r)\,
(f^2(r)-f^2(r_0))\,((\delta \acute x_2)^2+(\delta \acute x_3)^2)\phantom{\sum}\!\!\!\!\!\!\!\!\right]\,,
\label{lagr}
\eea
where $h^2(r)=g_{x}(r)g_{r}(r)$. The Euler-Lagrange equation for the $\delta x_1$ fluctuation is
\be
\left[\frac{d}{dr}\left(\frac{(f^2(r)-f^2(r_0))^{\frac32}}{g(r)f(r)}\frac{d}{dr}\right)+
\omega^2\frac{h^2(r) \sqrt{f^2(r) -f^2(r_0) }}{g(r)f(r)}\right]\delta x_1(r)=0\,,
\label{SL}
\ee
where we factorized the time dependence of the fluctuations as $\delta x(t,r)=\delta x(r)\,e^{-i\omega t}$.
The equations for the fluctuations transverse to the $(r,x_1)$-plane obtained from
(\ref{lagr}) are
\bea
\left[\frac{d}{dr}\left(\frac{f(r)\sqrt{f^2(r)-f^2(r_0)}}{g(r)}\frac{d}{dr}\right)+
\omega^2\frac{h^2(r)f(r)}{g(r)\sqrt{f^2(r)-f^2(r_0)}}\right]\delta x_m(r)=0\,,~~~~m=2,3\, .
\label{spp}
\eea
Equations (\ref{SL})-(\ref{spp}) are differential equations of the Sturm-Liouville type
defined for the half line $r_{min}\leq r_0 \leq r<\infty$ and we are interested in
analyzing the existence of instabilities, in particular
determining the range of values of $r_0$ for which $\omega^2<0$.

The boundary conditions to be
imposed on the problem are Dirichlet, this means fluctuations keeping the
string endpoints fixed at the boundary $\delta x_i(\tau,\sigma)|_{r=\infty}=0$,
but, the nature of the $r$-gauge parametrizing only half of the $(r_{\sf cl}(\sigma),x_{\sf cl}(\sigma))$
embedding requires an additional analysis of
boundary conditions at the tip $r=r_0$  (singular point of (\ref{SL})-(\ref{spp})). We start by analyzing (\ref{spp}), the
expansion around the tip gives\footnote{We consider $f(r)$ to be an increasing function of $r$ having no zeroes
except perhaps at the bottom of the bulk $r=r_{min}$.}
\be
\frac{d}{dr}\left(\!\sqrt{r-r_0}\;\frac{d\delta x_m(r)}{dr}\right)+
\frac{\omega^2h^2(r_0)}{2f(r_0)f'(r_0)}\frac1{\sqrt{r-r_0}}\,\delta x_m(r)\approx0~~~\Longrightarrow ~~~
\delta x_m(r)\approx C_0+C_1\sqrt{r-r_0}+O({r-r_0})  \,.
\label{xperp}
\ee
Here $C_{0,1}$ are arbitrary constants corresponding to the two independent  solutions of the differential equation (\ref{spp}), which once
chosen determine the whole series expansion for $\delta x_m(r)$. Physically they correspond respectively  to even and odd
fluctuations around the tip once we patch them with the fluctuations around the other half of the embedding,
which obviously satisfy the same eom. A discrete set of eigenvalues $\omega^2$ is expected if
non-normalizable solutions exist in the large
$r$-limit, we also expect the even solution to have the lowest $\omega^2$ eigenvalue (see appendix \ref{kmt}).

We now turn to the analysis of the in-plane $\delta x_1$
fluctuation. Expanding (\ref{SL}) around $r=r_0$ one finds
\be
\frac{d}{dr}\left((r-r_0)^{\frac32}\frac{d\delta
x_1(r)}{dr}\right)+\frac{\omega^2h^2(r_0)}{2f(r_0)f'(r_0)}\sqrt{r-r_0}\,\delta
x_1(r)\approx0 ~~~\Longrightarrow ~~~ \delta x_1(r)\approx
C'_0+C'_1\frac1{\sqrt{r-r_0}}+O(\sqrt{r-r_0}) \,.
\label{asymr0}
\ee
A singular behavior appears for $\delta x_1$ at the tip and one
might be tempted to cancel it by imposing $C'_1=0$. Nevertheless, as
mentioned above one should take notice that the $r$-gauge fixing
implies that, at the tip, $\delta x_1$  is directed along the string
worldsheet and not transverse to it, therefore in the $r$-gauge the
displacement $\delta x_1$ at the tip  is not physical. In order to
give a physical interpretation to the constants $C'_{0,1}$ in
(\ref{asymr0})   we now make a change from the $r$- to the
$x$-gauge (cf. \cite{gubser},\cite{avramis}).
The change of gauge on the ansatz (\ref{Avramis}) can be
implemented by a change of variables on the solution (\ref{SL}) from
$r$ to a new variable which we call $u$. It can be implemented
perturbatively, to first order in the
fluctuation the relation is \cite{avramis},
\be u=r+\Delta
(t,r)~~~~~\mathrm {with}~~~\Delta (t,r)=\frac{\delta
x_1(t,r)}{x'_{\sf cl}(r)}\, .
\label{change}
\ee
This transformation
performs the desired change of gauge since
\bea
x_1&=&x_{\sf cl}(r)+\delta x_1(t,r)=x_{\sf cl}(u-\Delta (t,r))+\delta
x_1(t,r)\nn\\&\approx& x_{\sf cl}(u)-x'_{\sf cl}(r)
\frac{\delta x_1(t,r)}{x'_{\sf cl}(r)}+\delta x_1(t,r)=x_{\sf cl}(u)\label{x1}\\
r&\approx&u-\frac{\delta x_1(t,u)}{x'_{\sf cl}(u)}\,,
\label{rr}
\eea
here $r_0\le u<\infty$. This is precisely the gauge fixing employed in \cite{pufu} mentioned above.
The second term on the rhs of (\ref{rr}) is
interpreted as the $r$-direction fluctuation induced by the ($r$-gauge) $x_1$-fluctuation.
It is now easy to see that it is finite. The asymptotic behavior (\ref{asymr0}) and
the expansion around the tip of (\ref{sig}), $x'_{\sf cl}(r)\sim(r-r_0)^{-1/2}$, plugged in (\ref{rr}) gives
\be
r\approx r_0- \alpha(C'_0{\sqrt{u-r_0}}+C'_1)+O(u-r_0)\,,
\label{rgauge}
\ee
with $\alpha$ a finite constant. The result (\ref{rgauge}) shows that the
physical $\delta r$ fluctuation originated from the non-physical divergent $\delta x_1$ fluctuation at the tip is
manifestly finite.
Therefore we interpret the ($C'_0$) $C'_1$    in (\ref{asymr0}) as the fluctuation that (do not)
oscillates the tip position.

\vspace{1.5mm}

In the following sections we will show numerical solutions of (\ref{SL}) for various
gravity backgrounds determining the lowest eigenvalues leading to normalizable solutions.
We will solve (\ref{SL}) by a shooting method integrating numerically from the tip $r_0$ up to a
large value $r_\infty$. The allowed values for $\omega^2$ will be obtained imposing the
numerical solution to be zero
at $r=r_\infty$. The boundary conditions at the tip corresponding to even solutions are
$C'_0=0$ and $C'_1$ arbitrary, for numerical purposes we set $C'_1=1$, its value sets the
normalization of the fluctuation. An even solution around the tip satisfies
\be
 \left.\frac{d\delta r(t,r)}{dx_1}\right|_{r=r_0}=0~~~~~\mathrm {where}~~~\delta r(t,r)=-\frac{\delta
x_1(t,r)}{x'_{\sf cl}(r)}\, .
\label{bcc}
\ee
Using (\ref{x1}) we can write (\ref{bcc}) in terms of $\delta x_1(r)$. The
boundary conditions for even solutions of (\ref{SL}) are implemented numerically as
\be
     \delta x_1(r)+2(r-r_0)\frac{d\delta x_1(r)}{dr}=0, ~~~~r\rightarrow
     r_0\nn
\ee
\be
~~~~~~~~~~~~~~~~~\sqrt{r-r_0}\,\delta x_1(r)=1\,
,~~~~r\rightarrow r_0\label{bc}\,.
\ee
Odd solutions $C'_0=1$ and $C'_1=0$ are implemented as
\be
     \delta x_1(r)+2(r-r_0)\frac{d\delta x_1(r)}{dr}=1,~~~~r\rightarrow
     r_0\nn
\ee
\be
~~~~~~~~~~~~~~~~~\sqrt{r-r_0}\,\delta x_1(r)=0\,
,~~~~r\rightarrow r_0\label{bc2}\,.
\ee
Summarizing, in general backgrounds, the functional relation of the classical solution
(\ref{sig}) between the $x_1$ and $r$ coordinates at the tip takes the form $x_{\sf cl}^2(r)\approx r-r_0$, and the asymptotic behavior of the
$x_1$-fluctuations over it, in the $r$-gauge, is of the form (\ref{asymr0}). Although a divergent
piece appears in (\ref{asymr0}), an appropriate change of gauge shows that the divergent and
non-divergent pieces correspond respectively to (physical) even and odd fluctuations around the tip.

\section{Gravity Backgrounds}
\label{bkg}

In this section we compute the string embeddings (\ref{eqxr}) dual to rectangular Wilson loops
for a a number of paradigmatic gravity backgrounds.
We review the $AdS_5\times S^5$ \cite{maldawilson} and $AdS_5$-Schwarzschild$\times S^5$
\cite{Brandsonnen} cases. Next we perform the numerical analysis of the equations (\ref{generalL})
and (\ref{enerren}) for the Maldacena-Nu{\~n}ez \cite{mn}, Klebanov-Strassler \cite{ks} and the
generalized Maldacena-Nu{\~n}ez \cite{cnp} backgrounds. In all cases the geometry is supported by
some non-trivial $p$-form fluxes, but they will not be relevant for our computations.

\subsection{$AdS_5\times S^5$}

This background is dual to $\cal N$ $=4$ SYM with $G=SU(N)$ gauge group in the Coulomb phase. The $AdS$ curvature $R$ relates to the gauge
theory 't Hooft coupling $\lambda$ as $R^4=\alpha'^2\lambda$ and the flux of the 5-form supporting the geometry $N=\int_{S^5} F_5$
relates to the rank of the gauge group as $N={\sf Rank}(G)$ \cite{malda}. The conformal invariance of the gauge theory
implies a Coulomb behavior for the potential $V(L)\sim 1/L$ between quarks. The novelty of the gravity computation
is to compute the gauge coupling dependence of the proportionality coefficient.

\subsubsection{Poincare coordinates  \cite{maldawilson}}

This coordinate system is supposed to describe the gauge theory formulated on $\mathbb R^{3,1}$. The metric is written as
\be
ds^2=\frac{r^2}{R^2}(-dt^2+dx_idx_i)+R^2\frac{dr^2}{r^2}+R^2d\Omega_5^2\,.
\label{ads5}
\ee
One finds $f^2(r)={r^4}/{R^4},~g^2(r)=1$. The radial coordinate range is $0<r<\infty$, at $r=0$ one finds a Killing horizon. Equation (\ref{sig}) can be analytically solved
in the $r$-gauge, one obtains
\be
x_{\sf cl}(r)=\pm\left\{ cte-\frac{R^2}{4r_0}\mathsf{B}\left(\left(\frac{r_0}r\right)^4;\frac34,\frac12\right)\right\},~~~~~~r_0\le r<\infty
\label{xads}
\ee
where $\mathsf{B}(z;a,b)$ is the incomplete beta function $\mathsf{B}(z;a,b)=\int_0^zt^{a-1}(1-t)^{b-1}dt$. The boundary conditions fix the constant
in (\ref{xads}) and relate the parameters
$r_0$ and $L$, setting $x_{\sf cl}(r_0)=0$ and $x_{\sf cl}(\infty)=\pm L/2$ one obtains \cite{maldawilson},
\be
L(r_0)=\frac{R^2}{2r_0}\mathsf{B}\left(\frac34,\frac12\right)=\frac{R^2}{r_0}\frac{(2\pi)^{\frac32}}{\Gamma[\frac14]^2}\,.
\label{lads}
\ee
The energy (\ref{enerren}) takes the form \cite{maldawilson}
\be
E_{q\bar q}(r_0)=\frac{r_0}{\pi\alpha'}(K(-1)-E(-1))=-\frac{r_0}{2\pi\alpha'}\frac{(2\pi)^{\frac32}}{\Gamma[\frac14]^2}\,,
\label{eads}
\ee
here $K(m),E(m)$ are the complete elliptic integrals of first and second kind.
Eliminating $r_0$ from expressions (\ref{lads})-(\ref{eads}) the AdS/CFT proposal for the interaction potential
between fundamental quarks in the large 't  Hooft coupling for the ${\cal N}=4$ SYM theory is \cite{maldawilson}
\be
V_{\mathsf{string}}(L)=-\frac{(2\pi)^{2}}{\Gamma[\frac14]^4}\frac{R^2/\alpha'}{L}\sim -\frac{\sqrt\lambda} L\,.
\label{Vads}
\ee
An attractive  Coulomb potential is obtained as expected from conformal invariance. The interesting result is the
$\sqrt\lambda=(g_{YM}^2N)^{\frac12}$ dependence when compared with the perturbative $\lambda=g_{YM}^2N$ result.
This suggests that some renormalization of the charges takes place at strong coupling \cite{maldawilson}. Note that
in order to obtain the negative quantity (\ref{Vads})
starting from a positive definite one (see eqn (\ref{energy})), it was crucial
to substract the quark masses (\ref{se}) .

\subsubsection{Global coordinates}
\label{adsglobal}

We discuss this example because it clarifies conceptual issues
regarding the interpretation of the substraction procedure
(\ref{enerren}) in smooth and complete gravity backgrounds (see
sections \ref{mnsol}, \ref{kssol}, \ref{gmnsol}).

\begin{figure}[h]
\hspace{-0.25cm}\centering
\begin{minipage}{13cm} 
\centering
\includegraphics[width=7.5cm]{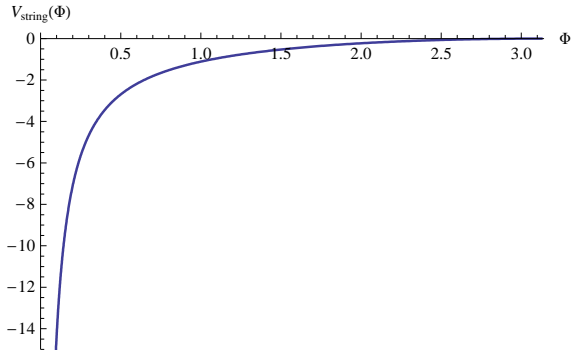}
\caption{$V_{\mathsf{string}}(\Phi)$ obtained from (\ref{fi})
and (\ref{enadsren}). For small angular
separations between the string endpoints $\Phi\ll1$ one finds the expected Coulomb behavior
$V\sim\sqrt\lambda/(R\Phi)$. For larger separations the solution deviates due to the compactness of the $S^3$.}
\label{Vadsglobal}
\end{minipage}
\end{figure}

Computations performed in $AdS$ global coordinates are supposed to represent the ${\cal N}=4$ SYM
gauge theory defined on $S^3\times \mathbb R$. The $AdS$ metric is now written as
\be
ds^2=R^2[-\cosh^2\!\rho\,dt^2+d\rho^2+\sinh^2\!\rho\, d\Omega_3^2]\,.
\ee
All coordinates are dimensionless in this case with the $AdS$ radius $R$ setting the scale. We write the
$S^3$ metric as $d\Omega_3^2=d\theta_1^2+\sin^2\!\theta_1(d\theta_2^2+\sin^2\!\theta_2d\varphi^2)$.
Being $\varphi$ a cyclic coordinate, the appropriate string ansatz is
$t=\tau,~\rho=\rho(\sigma),~\varphi=\sigma $, one obtains
$f^2(\rho)=\frac14\sinh^2\!2\rho,~g^2(\rho)=\cosh^2\!\rho$. The remaining angular variables must be set to
$\theta_i=\frac\pi2$ (equator of $S^3$) in order to satisfy their eom. The conserved charge in the $\varphi$
coordinate leads to an effective one dimensional zero energy motion
\be
\acute \rho^2+U(\rho)=0\,,
\label{global}
\ee
where $\acute \rho=d\rho/d\varphi$ and  the potential $U(\rho)=\sinh^2\!\rho\left(1-\frac{\sinh^2\!2\rho}{\sinh^2\!2\rho_0} \right)$. $\rho_0$
is the minimum radial position reached by the string when separating the string endpoints at infinity by
$\Delta\varphi=\Phi$. The $\Phi(\rho_0)$ relation (\ref{generalL}) is computed straightforwardly
\be
\Phi(\rho_0)=2\int_{r_0}^\infty \frac{\sinh 2\rho_0}{\sinh \rho\sqrt{\sinh^2\!2\rho-\sinh^2\!2\rho_0}}\,d\rho\,.
\label{fi}
\ee
Since the gauge theory is defined on a $S^3$ there exists a maximum separation for the quarks and it
corresponds to placing them
at antipodes on the equator of the $S^3$. This results in the string reaching the origin $\Phi(0)=\pi$
and leading to a smooth straight  worldsheet (stretched along the radial $r$ coordinate) parametrized by two
halves at $\varphi=\varphi_0$ and $\varphi=\pi+\varphi_0$ . The (divergent) energy (\ref{er})
of the configuration (\ref{global}) is
\be
E(\rho_0)=\frac R{2\pi\alpha'}\int_{\rho_0}^\infty  \frac{\sinh^2\!2\rho}{\sinh \rho \sqrt{\sinh^2\! 2\rho-\sinh^2\! 2\rho_0}}\,d\rho\,.
\label{diven}
\ee
Substracting the quark masses as in (\ref{enerren}) leads to,
\be
E_{q\bar q}(\rho_0)=\frac R{2\pi\alpha'}\left[\int_{\rho_0}^\infty \left(\frac{2\cosh \rho}{\sqrt{1-\frac{\sinh^2\!2\rho_0}{\sinh^2\!2\rho}}}
-2\cosh \rho\right)\,d\rho-2\sinh \rho_0\right]\,,
\label{enadsren}
\ee
which is finite and negative definite (see fig.\ref{Vadsglobal}). The finite result (\ref{enadsren}) should
be understood as resulting from the comparison of (\ref{diven}) wrt the aforementioned smooth reference state consisting in a straight string
with its endpoints at infinity at antipodes on the $S^3$ equator. We interpret this last configuration as the one
corresponding to  ``infinitely'' separated quarks on $S^3$. Note that the reference state and the configuration we are analyzing
satisfy different boundary conditions.

\subsection{$AdS_5$-Schwarzschild$\times S^5$}
\label{adssch}

\begin{figure}[h]
\begin{minipage}{7.5cm}
\begin{center}
\includegraphics[width=7.5cm]{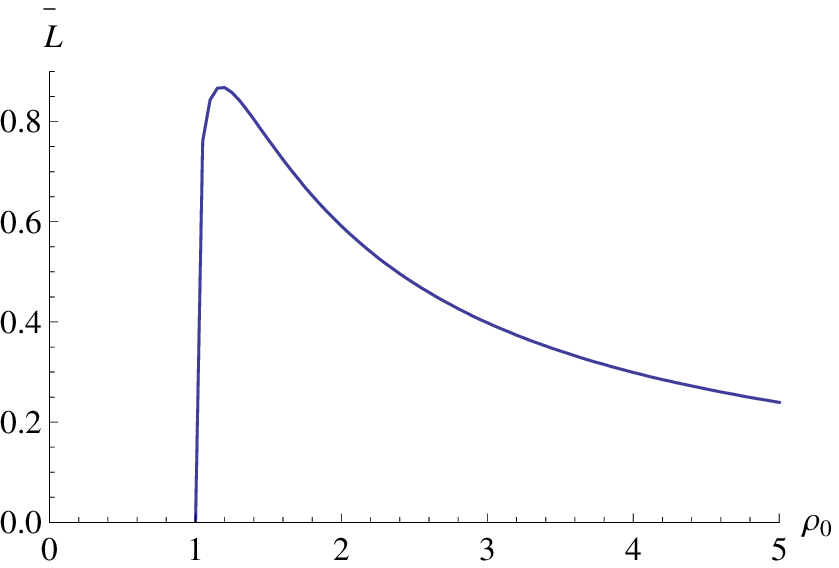}
\caption{$L(\rho_0)$ relation (\ref{ltads}): A maximum is observed for  $\rho_0\simeq1.177$. The
two branches to the left/right of the maximum at $\rho_{0c}\simeq1.177$ lead to a double valued $V_{\sf string} (L)$ relation in fig. \ref{E(L)}.}
\label{L(r_0)}
\end{center}
\end{minipage}
\   \
\hfill \begin{minipage}{7cm}
\begin{center}
\includegraphics[width=7.5cm]{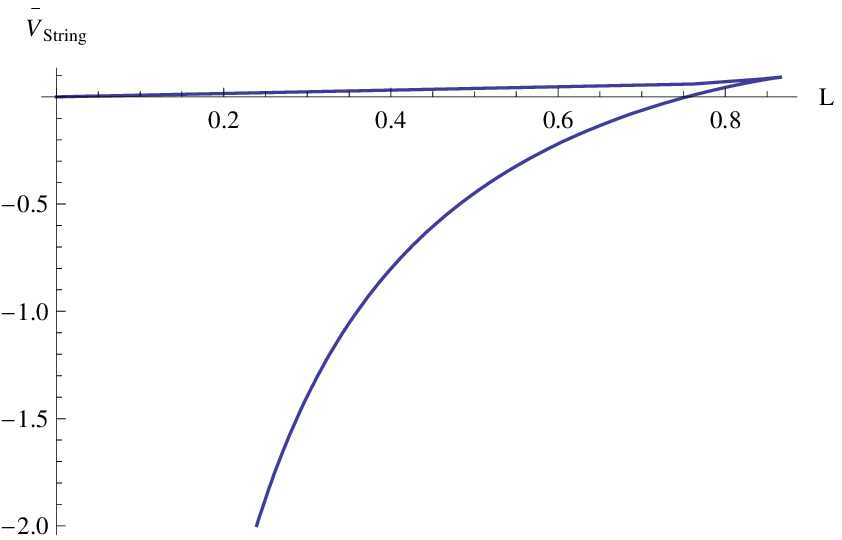}
\caption{Double valued $V_{\sf string}(L)$ obtained from (\ref{ltads}) and (\ref{etads})
eliminating numerically $\rho_0$. The upper curve
corresponding to the left branch in fig. \ref{L(r_0)} does not satisfy the conditions (\ref{convexity}).}
\label{E(L)}
\end{center}
\end{minipage}
\bc
\begin{minipage}{15cm}
\begin{center}
\includegraphics[width=7.5cm]{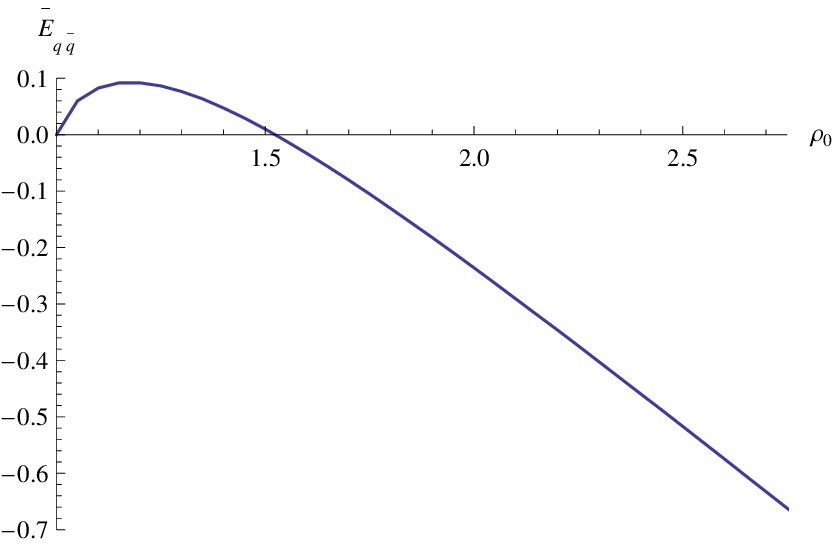}
\caption{$\bar E_{q\bar q}(\rho_0)$ relation for thermal ${\cal N}=4$ SYM. Eqn. (\ref{eprima}) guarantees
an extremum in $E_{q\bar q}(r_0)$ and $L(r_0)$ at the same value of $r_0$. The curve cuts the axis at $\rho_{0m}\simeq1.524$, for $\rho_0>\rho_{0m}$
the U-shaped solution is an absolute minimum, for $\rho_0<\rho_{0m}$ the two straight strings (reference) solution is the absolute minimum.}
\label{e(r_0)}
\end{center}
\end{minipage}
\ec
\end{figure}

Finite temperature gauge theories are described by considering black
hole (BH) solutions in the gravity duals \cite{wittenT}. The near
horizon geometry of $N$ non extremal (black) $D3$-branes is
therefore conjectured to describe ${\cal N}=4$ SYM at finite
temperature. As explained in \cite{wittenT}, the appropriate BH
background for describing the gauge theory on $\mathbb R^{3,1}$
involves a delicate infinite mass limit of the $AdS_5$-Schwarzschild
BH resulting in\footnote{The limit results in the metric depending
on only one scale, contrary to the finite mass case where one has
two parameters: the temperature (BH mass) and the $AdS$ radius. This
last geometry was shown to exhibit a phase transition (Hawking-Page),
which was interpreted as dual to the confiniement/deconfinement phase
transition in ${\cal N}=4$ on $S^3$ \cite{wittenT}.},
\be
ds^2=\frac{r^2}{R^2}\left[-(1-\frac{\mu^4}{r^4})dt^2+d{x_i}dx_i\right]
+\frac{R^2}{r^2}\frac1{1-\frac{\mu^4}{r^4}}dr^2+R^2d\Omega_5^2\,.
\label{thermalads}
\ee
The BH horizon is located at $r=\mu$ and its
temperature is $T=\frac\mu{R^2}\pi$. It is convenient to work with
dimensionless coordinates, scaling $r=\mu\,\rho$, $t=R^2/\mu\, \bar
t$ and $x=R^2/\mu\, y$ one obtains
\be
ds^2=R^2\left[-\left({\rho^2}-\frac{1}{\rho^2}\right)d\bar
t^2+{\rho^2}d{y_i}dy_i+\frac1{{\rho^2}-\frac{1}{\rho^2}}d\rho^2+d\Omega_5^2\right].
\label{adimTads}
\ee
The scale of the dimensionless gauge theory
coordinates $\bar t,y_i$ in (\ref{adimTads}) is set by $R^2/\mu$ and
one finds $f^2(\rho)=\rho^4-1,~g^2(\rho)=1$ and $\rho=1$ as the
horizon location. The expressions for the dimensionless $q\bar{q}$
separation length (\ref{generalL}) and potential energy
(\ref{enerren}) can be analytically computed
\cite{avramis,Brandsonnen,Brandh}
\be \bar
L(\rho_0)=\frac{(2\pi)^{\frac32}}{\Gamma[\frac14]^2}\frac{\sqrt{\rho_0^4-1}}{\rho_0^3}\,\,_2F_1\!
\left(\frac34,\frac12,\frac54;\frac{1}{\rho_0^4}\right)
\label{ltads}
\ee
\be \bar
E_{q\bar{q}}(\rho_0)=\frac{R^2}{\pi\alpha'}\left[1-
\frac{(2\pi)^{\frac32}}{2\Gamma[\frac14]^2}\,\rho_0
\,_2F_1\!\left(-\frac12,-\frac14,\frac14;\frac{1}{\rho_0^4}\right)\right]
\label{etads}
\ee
Here $\rho_0\ge1$ is the minimum radial position
reached by the string and the minimum radial value $r_{min}$ in
(\ref{enerren}) was taken to be the horizon  location
 $r_{min}=\mu$.  One can easily check
that in the small temperature limit $LT\ll1$ (corresponding to
$\rho_0\gg1$) one recovers the zero temperature behavior
(\ref{lads})-(\ref{Vads}). We have plotted in figures \ref{L(r_0)}
and \ref{e(r_0)} the behavior of the length (\ref{ltads}) and the
energy (\ref{etads}) as functions of $\rho_0$. In figure \ref{E(L)}
we plotted the relation $V_{\sf string}(L)$ obtained from
(\ref{ltads})-(\ref{etads}) by eliminating $\rho_0$, the result is a double valued
function.

Figure \ref{L(r_0)}
shows  a maximum $\bar L_c\simeq0.869$ at
$\rho_{0c}\simeq1.177$ which implies that no smooth solution connecting the  pair of quarks
exists for $\bar L>\bar L_c$. The only existing solution for $\bar L>\bar L_c$ corresponds to two straight strings reaching the
horizon. This configuration, used for the substraction in (\ref{enerren}), is
interpreted as the one corresponding to a pair of free quarks. The existence of a maximum in the $L(r_0)$
relation in BH backgrounds has been interpreted
as the gravity dual of thermal bath screening  \cite{Brandsonnen}.
Figure \ref{L(r_0)} also shows the existence of two branches of solutions for each string endpoints separation
distance $L<L_c$.  The left branch ($L'>0$) leads to a potential $V_{\sf string}(L)$ not satisfying
the conditions (\ref{convexity}) and it should therefore be non-physical (upper curve in fig. \ref{E(L)}).
The rewarding result as we shall see is that string theory is wise: the left branch should not be trusted since it is unstable
under small perturbations (see section \ref{nond3}).

The last point to comment is that although, as we shall show, all the $L'<0$ region ($1.177<\rho_0<\infty $)
is stable under small perturbations,
one expects the lower curve in figure \ref{E(L)}
solution to be metastable whenever $E_{q\bar q}>0$ ($0.754<\bar L<0.869$ or $1.177<\rho_0<1.524$).
The reason for this
is that  the (reference) two straight lines solution has $E_{q\bar q}=0$  and is therefore the absolute
stable minimum fro $0.754<\bar L<0.869$ (see fig. \ref{e(r_0)}). Taking this last fact into account  \cite{Brandsonnen} suggested to
take $\bar L_{max}\simeq0.754$ as the screening length.

\subsection{Maldacena-N\'u\~nez  background \cite{mn}}
\label{mnsol}

  The $r\approx0$ of this background is supposed to describe qualitatively the IR regime of $d=4$
$\cal N$ $=1$ SYM theory. The probe brane configuration leading to
this solution consists of $N$ $D5$-branes wrapped on a finite
2-cycle at the origin of the resolved conifold. When the
backreaction of the $D5$ on the geometry is taken into account, a
transition flop occurs (see \cite{vafa}) leading to a geometry  with
a smoothly collapsing $S^2$ and a finite $S^3$ at the origin as for
the deformed conifold (see \cite{paredes} for a review of the
solution). The solution was independently found in the context of
gauged supergravity in \cite{chv}.

The metric can be written as \cite{mn}
\beq
ds^2= \alpha'  Ne^{{\phi}}\,\,\Big[-dt^2+dx_idx_i+
dr^2+e^{2h}\,(d\theta^2+\sin^2\theta d\varphi^2)+ {\frac1
4}\,(w^i-A^i)^2\Big]\,, \label{metric}
\eeq
where $w^i$ ($i=1,2,3$) are the $su(2)$ right-invariant forms
\ba
 \lab{su2}
 w^1+iw^2= e^{-i\psi}(d\tilde\theta+i\sin\tilde\theta\, d\tilde\varphi),~~~~
 w^3=d\psi\,+\,\cos\tilde\theta\, d\tilde\varphi\,,
\ea
and $A^i,\phi, h$  are given by
\ba
A^1&=&-a(r)\, d\theta\,, \,\,\,\,\,\,\,\,\, A^2\,=\,a(r) \sin\theta\,
d\varphi\,, \,\,\,\,\,\,\,\,\, A^3\,=\,- \cos\theta\, d\varphi\,\nn\\
 a(r)&=&\frac{2r}{\sinh 2r}\nn\\
 e^{2h}&=&r\coth 2r\,-\,\frac{r^2}{\sinh^2 2r}\,-\, {\frac1 4}\nn\\
e^{2\phi}&=&e^{2\phi_0}\, \frac{\sinh 2r}{ 2e^h}\,.
\label{oneform}
\ea
$\phi_0$ is an integration constant which sets the value of the dilaton $\phi$ at the origin
(some authors write $g_s=e^{\phi_0}$).
The $t,x_i,r$ coordinates are dimensionless and its scale is set by  $(\alpha'  N)^{\frac12}$.
The metric (\ref{metric}) is regular in the $r\to 0$ limit with the 2-sphere ($\theta,\varphi$) smoothly
collapsing and the resulting topology of the
spacetime is of the form ${\cal M}_7\times S^3$, contrary to the cases discussed before
where the spacetimes were of the form ${\cal M}_5\times X^5$ with $X^5$ compact. The reason for
this is that the background (\ref{metric}) models the 5+1 gauge theory on the wrapped $D5$-branes.
Nevertheless, one expects to get an effective 3+1 theory at energies $E<1/R_{S^2}$, where
$R_{S^2}$ is the radius of the sphere wrapped by the $D5$.

\begin{figure}[h]
\begin{minipage}{7cm}
\begin{center}
\includegraphics[width=7.5cm]{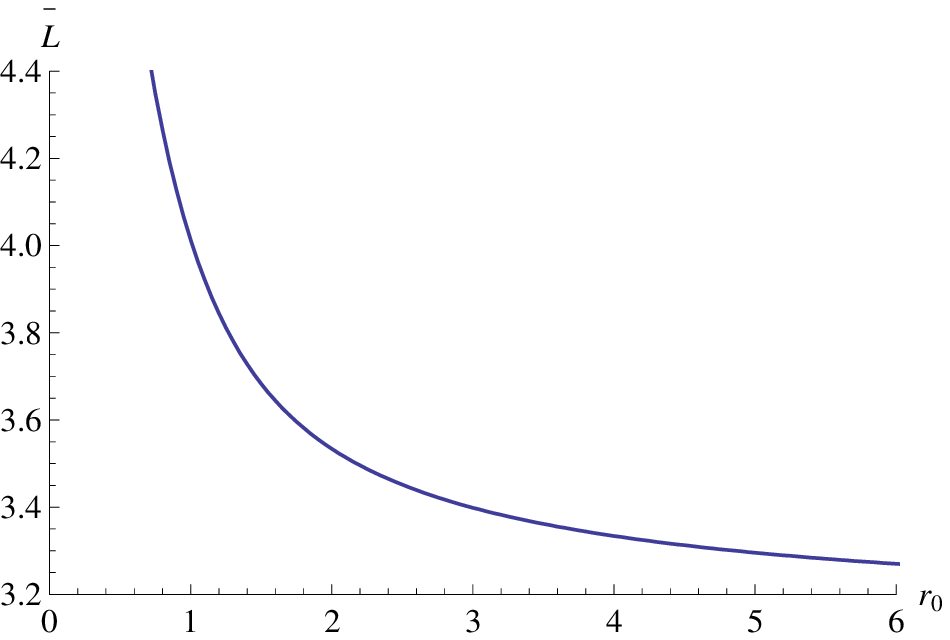}
\caption{$\bar L(r_0)$ relation (\ref{lmn}). MN solution.}
\label{lvsRmin MN}
\end{center}
\end{minipage}
\   \
\hfill \begin{minipage}{7cm}
\begin{center}
\includegraphics[width=7.5cm]{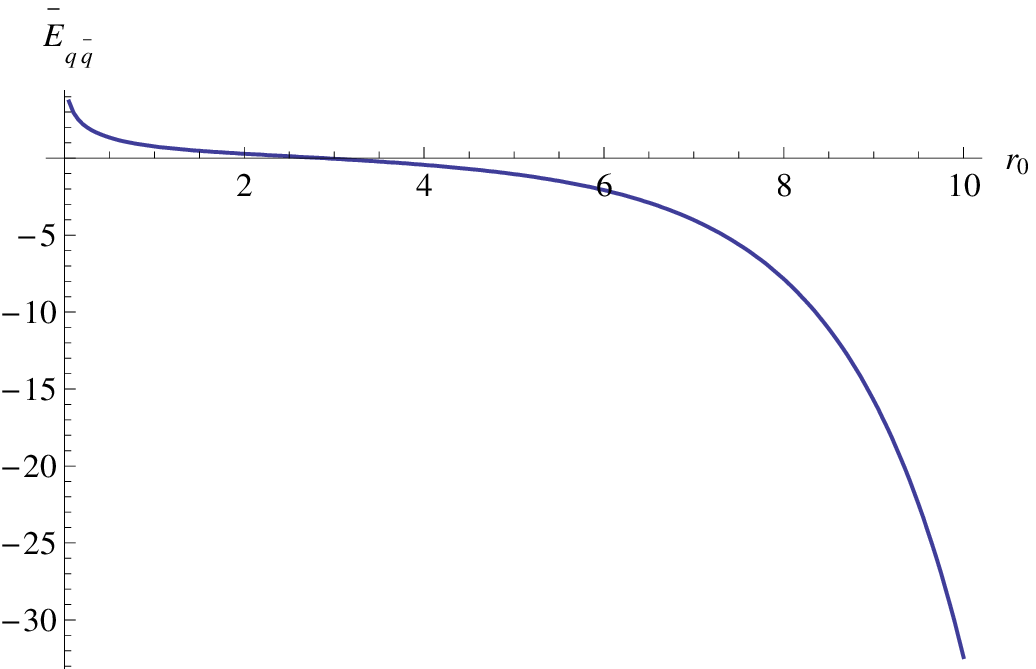}
\caption{$E(r_0)$ relation (\ref{emn}). MN solution.}
\label{EvsRmin MN}
\end{center}
\end{minipage}
\end{figure}

The paradigmatic static U-shaped ansatz leads to $f^2(r)=g^2(r)=e^{2\phi}$ and the separation length
(\ref{generalL}) and energy (\ref{enerren}) take the form
\be
\bar L(r_0)=2\int_{r_{0}}^{\infty}{\frac{e^{\phi(r_{0})}}{\sqrt{e^{2\phi(r)}-e^{2\phi(r_{0})}}}}\,dr
\label{lmn}
\ee
\be
\bar E_{q\bar q}(r_0)=\frac {N }{\pi}\left[\int_{r_{0}}^{\infty}\frac{e^{2\phi(r)}}{\sqrt{e^{2\phi(r)}-e^{2\phi(r_{0})}}}\,dr-
\int_{0}^{\infty}e^{\phi(r)}dr\right]\,.
\label{emn}
\ee
The expression for the energy (\ref{emn}) can be rewritten in an illuminating form \cite{sonnen}
\bea
\bar E_{q\bar q}(r_0)&=&\frac {N}{\pi}\left[\int_{r_{0}}^{\infty}\left(\frac{e^{2\phi(r)}+e^{2\phi(r_0)}-e^{2\phi(r_0)}}
{\sqrt{e^{2\phi(r)}-e^{2\phi(r_{0})}}}-e^{\phi(r)}\right)\,dr-
\int_{0}^{r_0}e^{\phi(r)}dr\right]\nn\\
&=&\frac {N}{\pi}\left[e^{\phi(r_{0})}\frac{\bar L(r_0)}2+\int_{r_{0}}^{\infty}{dr(\sqrt{e^{2\phi(r)}-e^{2\phi(r_{0})}}-e^{\phi(r)})}
 -\int_0^{r_{0}}{e^{\phi(r)}}dr\right]\,.
 \label{energMN}
\eea
In the large $L$ limit the string reaches the bottom of the bulk
($r_0\to0$, see figure \ref{lvsRmin MN}) and the last two terms in (\ref{energMN})
do not contribute (see \cite{sonnen}). In the large $\bar L\gg1$   limit we obtain reinserting units   \cite{mn}
\be
V_{\sf string}(L)\approx\frac {e^{\phi_0}}{2\pi\alpha'} L \quad\Rightarrow\quad T_{\sf string}
=\frac {e^{\phi_0}}{2\pi\alpha'}\,.
\label{tension}
\ee
The background (\ref{metric}) is therefore predicting linear confinement for large quarks separation in agreement with its proposed
$d=4$ ${\cal N}=1$ SYM dual gauge theory. From the (chromoelectric) string tension (\ref{tension}) we see that
the value of the dilaton at the origin $\phi_0$ relates to the dynamically generated scale of the dual gauge theory.
\\
We plot in figures \ref{lvsRmin MN} and \ref{EvsRmin MN} the $\bar L(r_0)$ and $\bar E(r_0)$ relations (\ref{lmn}) and (\ref{emn}). The
divergence $L\to \infty$ at $r_0\to 0$ in fig. \ref{lvsRmin MN} is due to $\left.\frac d{dr}(e^{2\phi(r)})\right|_{r=0}=0$ (see \cite{piai}
for a recent discussion).
\begin{figure}[h]
\bc
\begin{minipage}{15cm}
\begin{center}\includegraphics[width=7.5cm]{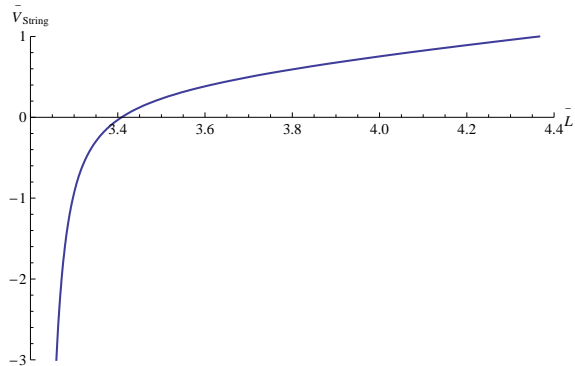}
\caption{Potential energy $V_{\sf string}(L)$ for the $q\bar q$ pair
obtained for the Maldacena-Nu{\~n}ez solution eliminating numerically
$r_0$ from (\ref{lmn})-(\ref{emn}). Note the change of behavior from
Coulomb like to linear.}
\label{EwilsonMN}
\end{center}
\end{minipage}
\ec
\end{figure}\\

The result for the potential $V_{\sf string}(L)$ in figure \ref{EwilsonMN} is rewarding, but the linear behavior
occurs for configurations having energies above zero. A concern arises as to whether we should trust
the result for $V_{\sf string}>0$ (see last paragraph of section \ref{adssch}). The substraction in
(\ref{emn}) corresponds to a pair of straight strings running along the radial direction with the remaining
spatial coordinates kept fixed. Being the background regular, the strings cannot end at any point in the interior
and the only possibility for a smooth reference solution is to place the string endpoints at antipodes on the $\varphi$
coordinate and having both the same $x_1$ coordinate (see the horizontal blue line in fig. \ref{substraction} and the related discussion
in sect. \ref{adsglobal}). We conclude that although the linear confinement occurs for configurations having
$E_{q\bar q}>0$, the solution is stable and it cannot decay to the reference $E_{q\bar q}=0$ state since the two
configurations being compared in (\ref{emn}) satisfy different boundary conditions. Figure \ref{substraction}
depicts the relevant worldsheets for the rectangular Wilson loop computations in smooth gravity backgrounds.

\vspace{1cm}

\begin{figure}[h]
\begin{center}
\hspace{-2cm}
\psline[linewidth=.3pt]{->}(1,-.7)(1,0.3)
\psline[linewidth=.3pt]{->}(1,-.7)(2,-.7)
\rput[bI](0.8,.5){$x_1$}
\rput[bI](2.2,-.8){$r$}
\psbezier[linewidth=.3pt]{->}(.7,-.2)(.8,-.4)(1.2,-.4)(1.3,-.2)
\rput[bI](1.6,-.3){$\varphi$}
\psellipse[linewidth=.5pt](1,-1.5)(2,.3)
\psbezier[linewidth=.5pt](-1,-6)(-.90,-6.5)(2.9,-6.5)(3,-6)
\psbezier[linewidth=.5pt,linestyle=dashed](-1,-6)(-.90,-5.5)(2.9,-5.5)(3,-6)
\psline[linewidth=.5pt](-1,-6.)(-1,-1.5)
\psline(-1,-6)(3,-6)
\psline[linecolor=blue,showpoints=true](-1,-6)(3,-6)
\psline[linewidth=.5pt](3,-6.)(3,-1.5)
\psline[linewidth=.3pt]{<->}(3.3,-6.)(3.3,-2)
\rput[bI](4,-4){$L$}
\psline[linewidth=.3pt,linestyle=dashed](1,-1)(1,-6.7)
\pscurve(3,-6)(1.4,-5)(1.3,-4)(1.4,-3)(3,-2)
\psdots[dotsize=0.2](3,-2)(-1,-6)(3,-6)
\vspace{7cm}
\caption{Geodesics employed for the computation of rectangular Wilson loops in smooth gravity
backgrounds. The black curved vertical line
depicts the U-shaped solution, the horizontal blue line is the smooth reference state with respect
to which we compare the energy of the black line. The black and blue configurations satisfy different boundary conditions.}
\label{substraction}
\end{center}
\end{figure}

\subsection{Klebanov-Strassler background \cite{ks}}
\label{kssol}

\begin{figure}[h]
\begin{minipage}{7cm} 
\bc
\includegraphics[width=7.5cm]{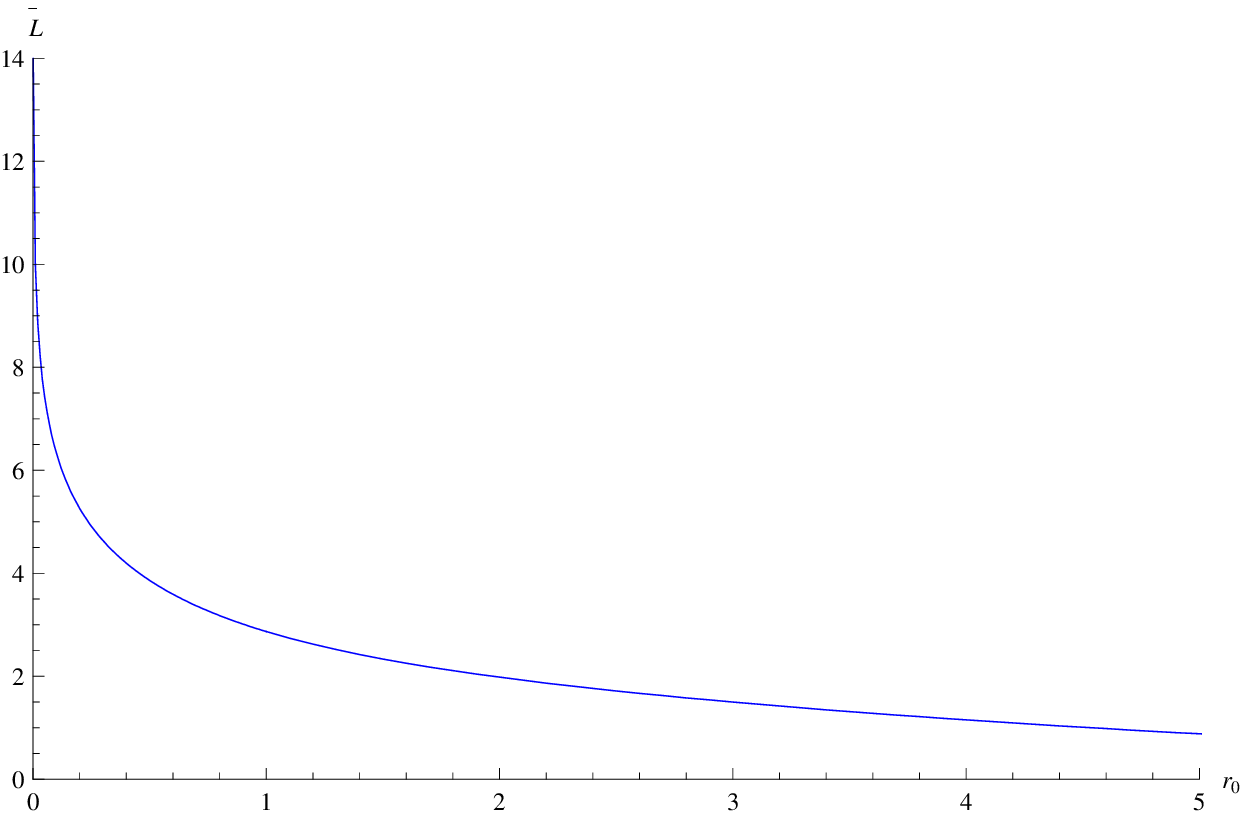}
\caption{$\bar L(r_0)$ relation (\ref{lks}). KS solution.}
\label{LvsrminKS}
\ec
\end{minipage}
\   \
\hfill
\begin{minipage}{7cm}
\begin{center}
\includegraphics[width=7.5cm]{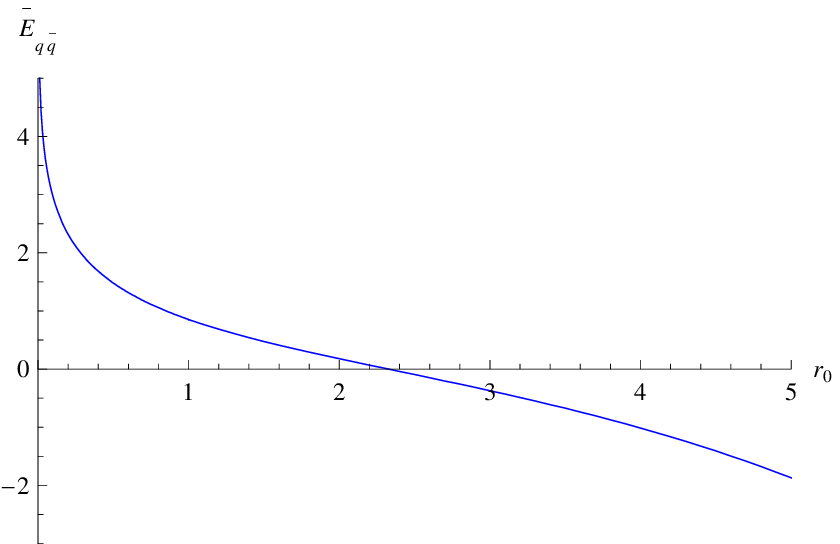}
\caption{$\bar E_{q\bar q}(r_0)$ relation (\ref{eks}). KS solution.}
\label{EvsRminKS}
\end{center}
\end{minipage}
\end{figure}

This background describes a $\cal N$ $=1$
quiver gauge theory with bifundamental matter fields transforming under $SU(N+M)\times SU(N)$.
The probe branes configuration leading to this geometry is constructed as:  $N~D3$-branes in the apex of the singular
conifold plus $M~D5$-branes wrapped in the topological $S^2$ of the conifold and sharing the remaining three dimensions
with the $D3$. The solution is supported by a  constant dilaton $\phi=\phi_0$, which can be set
to have $g_s\ll1$ everywhere (contrary to the MN solution). The $\cal N$ $=1$ flow $SU(N+M)\times SU(N)\rightarrow SU(M)$
realized through a cascade of Seiberg dualities in the gauge theory (see \cite{Strassler}) manifests in the geometry
in a varying 5-form flux.

\begin{figure}[h]
\bc
 \includegraphics[width=7.5cm]{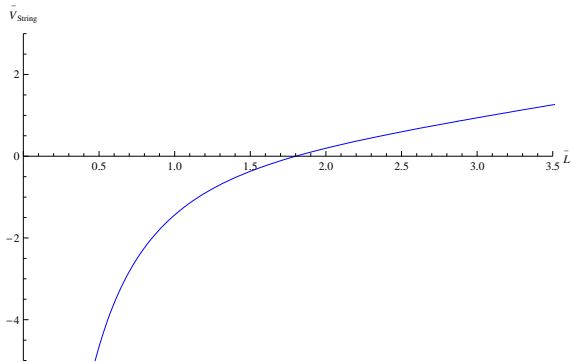}
\caption{$V_{\sf string}(L)$ relation for a rectangular Wilson loop in the KS solution.}
\label{EwilsonKS}
 \ec
\end{figure}

The metric reads \cite{ks}
\be
ds^2=g_s \alpha'\! M\, [h^{-\frac12}(r)(-dt^2+dx_{i}dx_{i})+h^{\frac12}(r)ds_{6}^2] \label{ks}
\ee
The deformed conifold metric $ds_6$ can be written
\bea ds_6^2& =& \frac{1}{2} K(r)\left[\frac{(dr^2 +
(g^5)^2) }{3 K^3(r)} +
\cosh^2\frac{r}{2}\, ((g^3)^2 +(g^4)^2) + \sinh^2 \frac{r}{2} \,
((g^1)^2 + (g^2)^2) \right],
\label{defconif}
\eea
where
\be
K(r)=\frac{[\sinh (2r) - 2r]^{\frac13}}{2^{\frac13}\sinh r}
\label{kks}
\ee
and the $g_i$  defined by
\bea
g^1 = \frac{e^1 - e^3}{\sqrt2}, \quad g^2 = \frac{e^2 -
e^4}{\sqrt2},\quad g^3 = \frac{e^1 + e^3}{\sqrt2}, \quad g^4 =
\frac{e^2 + e^4}{\sqrt2},\quad g^5 = e^5, \label{fbasis}
\eea
with
\begin{eqnarray}
e^1= -\sin\theta_1\,d\phi_1, ~~~e^2= d\theta_1, \qquad
e^3= - \sin\psi\, d\theta_2+\cos\psi\,\sin\theta_2\,d\phi_2 , \nonumber \\
e^4=\cos\psi\, d\theta_2+\sin\psi\,\sin\theta_2\,d\phi_2 , \qquad e^5=
d\psi + \cos\theta_1\,d\phi_1 + \cos\theta_2\,d\phi_2.
\end{eqnarray}
The coordinates in (\ref{ks}) are dimesionless, the gauge theory coordinates $t,x_i$  scale is set
by $\frac{g_s M \alpha'}{\ell_{cf}}$, and the scale of the holographic coordinate $r$ is set by $\ell_{cf}$.
The $h(r)$ factor takes the form
\bea
h(r) = 2^{\frac23}
\int_r^\infty d x \frac{x\coth x-1}{
\sinh^2 x}\, (\sinh 2x - 2x)^{\frac13}.
\label{hks}
\eea
The background (\ref{ks}) is supported in part by a non-trivial $B_{\mu\nu}$
but the embedding we are considering gets no contribution from it \cite{loewy}.
The functions in (\ref{geff}) are given by $f^2(r)=\frac{1}{h(r)},~g^2(r)=\frac{1}{6K^2(r)}$.
The dimensionless expressions for
the length (\ref{generalL}) and the energy (\ref{enerren}) are
\be
\bar {L}(r_0)=2\int_{r_0}^\infty\frac{dr}{\sqrt6K(r)}\frac{{h(r)}}{\sqrt{h(r_0)-h(r)}}\,.
\label{lks}
\ee
\be
\bar{E}_{q\bar q}(r_0)=\frac{g_sM}{\pi }\left[
\int_{r_0}^\infty\frac{dr}{\sqrt6K(r)}\frac{\sqrt{h(r_0)}}{\sqrt{h(r_0)-h(r)}}-\int_0^{r_0}\frac{dr}{\sqrt6K(r)}\right]\,.
\label{eks}
\ee
In figures \ref{LvsrminKS} and \ref{EvsRminKS} we have plotted these two last expressions. As in the MN case a divergence is
expected at $r_0=0$ since $\left.\frac {dh(r)}{dr}\right|_{r=0}=0$ Eliminating numerically $r_0$
from (\ref{lks})-(\ref{eks}) we plot in figure \ref{EwilsonKS} the $V_{\sf string}(L)$ function. A linear relation for the
interaction potential is observed for $L\gg1$. Proceeding as in (\ref{energMN}) we find the confining string tension to be
\be
T_{\sf string}=\frac{1}{2\pi\alpha'}\frac{\ell_{cf}^2}{g_s \alpha'\! M\sqrt{h_{0}}}\,,
\label{ts}
\ee
where $h_0=h(0)\simeq1.1398$. As in the Maldacena-N\'u\~nez case, the dominant contribution to the
minimal area (\ref{NG}), in the large $L$ limit, comes from the $r\approx0$ region. Again,
a concern arises regarding whether one should trust the $E_{q\bar q}>0$ configurations, as discussed at the
end of the last section
and pictured in fig. \ref{substraction} the $E_{q\bar q}>0$ embeddings are classically stable.

\subsection{Generalized Maldacena-N\'u\~nez solutions \cite{cnp},\cite{hn}}
\label{gmnsol}

This class of backgrounds was obtained  in \cite{cnp} generalizing the solution
described in section \ref{mnsol}. The solutions
were thoroughly discussed in  \cite{hn} an interpreted as dual to minimally
supersymmetric gauge theories
containing irrelevant dimension six operators. The operator drastically changes the
UV behavior of the theories taking the solution `away' from the near horizon of the
$D5$-branes which generate the geometry. The analysis in \cite{hn} shows that the
solutions asymptote, for large $r$, four dimensional Minkowski times the deformed conifold.

\begin{figure}[h]
\begin{minipage}{7cm}
\begin{center}
\includegraphics[width=7.5cm]{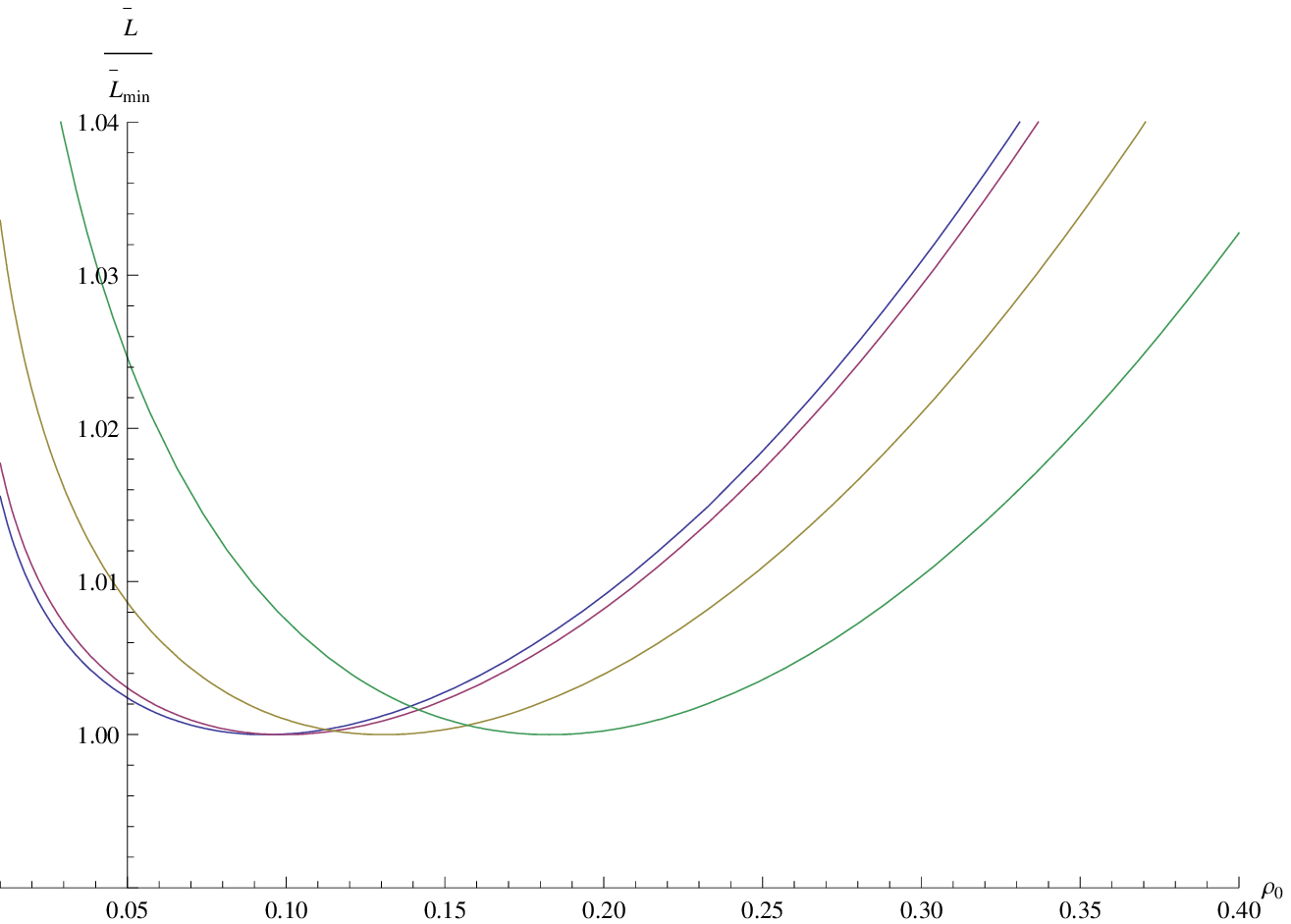}
\caption{$L(\rho_0)$ relation (\ref{LMNgen}). Generalized MN background with blue, violet, yellow and
green curves corresponding to
$\mu=-1.8,\,-1.5,\, -1,\,-.8$
($\rho_\infty=7$). A minimum quark separation length is observed.}
\label{LMNgmumenosuno}
\end{center}
\end{minipage}
\   \
\hfill \begin{minipage}{7cm}
\begin{center}
\includegraphics[width=7.5cm]{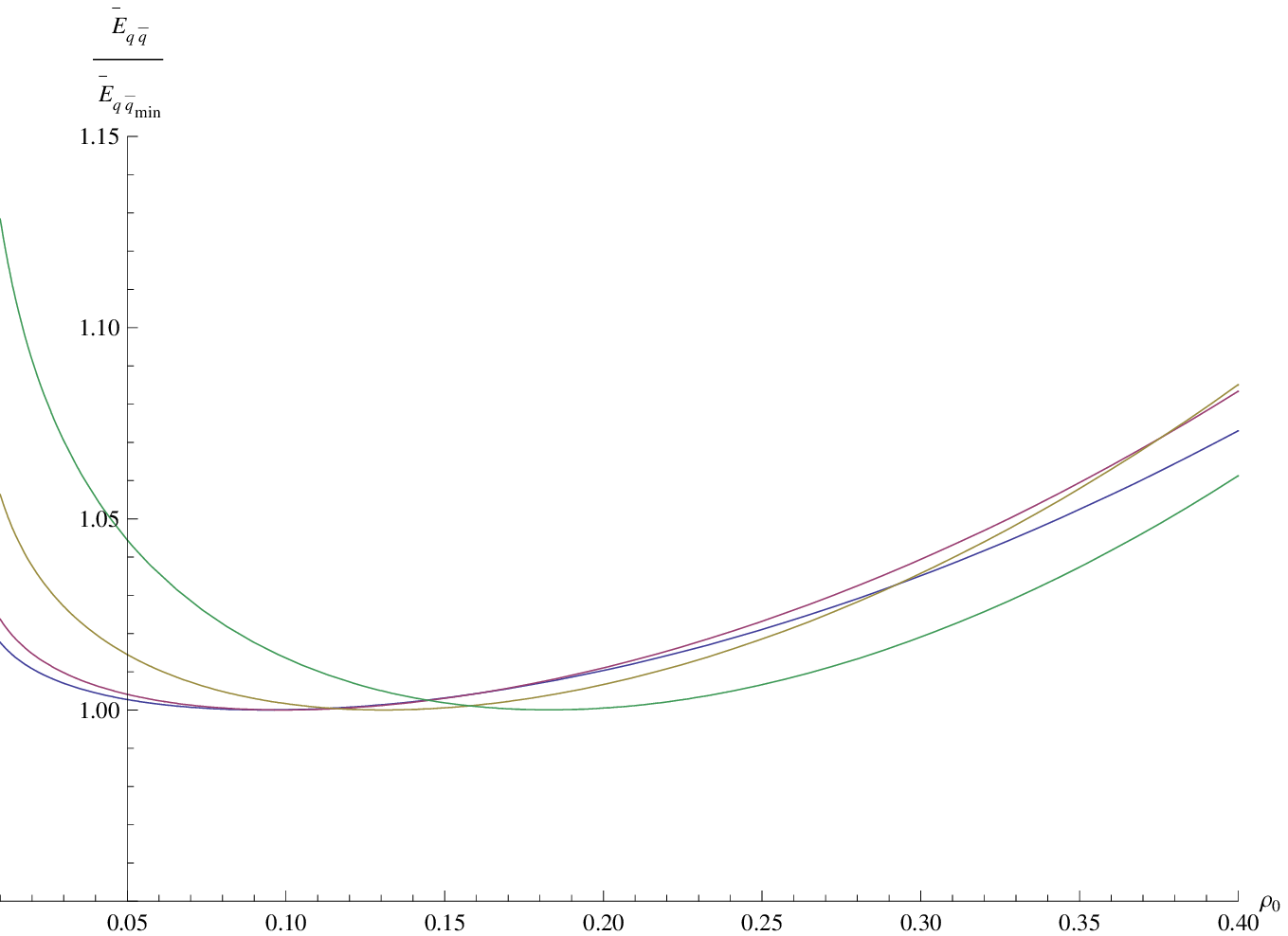}
\caption{$E(r_0)$ relation (\ref{EMNgen}). gMN background colors as in fig. \ref{LMNgmumenosuno}.}
\label{EMNgmumenosuno}
\end{center}
\end{minipage}
\end{figure}

The generalized MN metric reads \cite{cnp}
\bea
ds^2&=&g_s \alpha'\! N\, e^{{4f(r)}}  \Big[-dt^2+dx_{i}dx_{i}+ dr^2+e^{2h(r)}\,(d\theta^2+\sin^2\theta d\varphi^2)\nn\\
&&+ \frac{e^{2g(r)}}{4}\left((w_1+a(r)d\theta)^2+(w_2-a(r)\sin\theta d\varphi)^2\right)
+\frac{e^{2k(r)}}{4}(w_3+\cos\theta d\varphi)^2\Big]\,.
\label{gMN}
\eea
Making the change of coordinates $d\rho\equiv e^{-k(r)}dr$, the functions $a$, $k$ and $f$ satisfy
\bea
\partial_\rho a &=&\frac{-2}{-1+2\rho\coth2\rho}\left[{e^{2k}} \frac{(a\cosh2\rho-1)^2}{\sinh2\rho}+a\,(2\rho
-a\sinh2\rho)\right]\nn\\
\partial_\rho k &=&\frac{2(1+a^2-2a\cosh2\rho)^{-1}}{-1+2\rho\coth2\rho}\left[ {e^{2k}} a\sinh2\rho\,(
a\cosh2\rho-1)+(2\rho-4a\,\rho\cosh2\rho+\frac{a^2}{2}\sinh4\rho)\right]\nn\\
\partial_\rho f &=&-\frac1{4\sinh^{2}2\rho}\left[\frac{(1-a \cosh 2\rho )^2(-4\rho+\sinh4\rho)}{(1+a^2-2a\cosh2\rho)(-1+2\rho\coth2\rho)}\right],
\label{eqdf}
\eea
and the  $g(\rho),h(\rho)$ functions in (\ref{gMN}) are given by
\be
e^{2g }=\frac{b \cosh 2\rho -1}{a \cosh 2\rho -1},~~~~
e^{2h }=\frac{e^{2g }}{4}(2a \cosh 2\rho -1-a^2),~~~\mathrm{with}~~b(\rho)=\frac{2\rho}{\sinh2\rho}\,.\label{bghMNg}
\ee
The first two differential equations in (\ref{eqdf}) have a one parameter family of regular solutions. For small $r$ one finds \cite{cnp}
\be
a(\rho)=1+\mu\rho^2+...,~~~~e^{2k(\rho)}= \frac{4}{6+3\mu}-\frac{20+36\mu+9\mu^2}{15(2+\mu)}\rho^2+....
\label{akMN}
\ee
with $\mu$ taking values in the interval $(-2,-\frac23)$. Inserting (\ref{akMN}) into the third equation of (\ref{eqdf}) and into (\ref{bghMNg})
one obtains
\be
e^{2g(\rho)}= \frac{4}{6+3\mu}+...,~~~~e^{2h(\rho)}= \frac{4\rho^2}{6+3\mu}+....,~~~~e^{2f(\rho)}=1+\frac{(2+\mu)^2}{8}\rho^2+...
\ee
The arbitrary constant for $f$ following from (\ref{eqdf}) was factored out as $g_s$ in (\ref{gMN}).
The limit values for $\mu$ give known solutions: the $\mu=-\frac23$ case reproduces the MN solution of
section \ref{mnsol} with $\phi=4f$  ($k=const.$), and the case $\mu=-2$ case leads to 4-dimensional Minkowski space times the deformed conifold
($\phi$ being constant in this case).
Finally, the $\rho\rightarrow\infty$ limit of all solutions (except $\mu=-\frac23$) asymptotes the deformed conifold metric
(see  \cite{cnp} for details).

The length (\ref{generalL}) and energy (\ref{enerren}) expressions  are given by
\be
\bar
L(\rho_0)=2\int_{\rho_0}^{\rho_\infty}\frac{e^{4f(\rho_0)}}{\sqrt{e^{8f(\rho)}-e^{8f(\rho_0)}}}\,e^{k(\rho)}d\rho
\label{LMNgen}
\ee
\be
\bar
E_{q\bar{q}}(\rho_0)=\frac{Ng_s}\pi\left[\int_{\rho_0}^{\rho_\infty}
\frac{e^{8f(\rho)}}{\sqrt{e^{8f(\rho)}-e^{8f(\rho_0)}} }\,
e^{k(\rho)}d\rho -\int_{0}^{\rho_\infty}
e^{4f(\rho)}e^{k(\rho)}d\rho\right]\,.
\label{EMNgen}
\ee
 Note that the radial integration in both expressions extends up to a finite
distance $\rho_\infty$. The reason, noted in \cite{cnp}, being that
(\ref{LMNgen}) is UV divergent\footnote{See \cite{piai} for a discussion on divergences
when computing the length function $L(r_0)$.}. We  should therefore have in mind
that the string computation corresponds to probing  the dual gauge theory with very massive
for $\rho_\infty\gg1$ (but not infinite mass)  quarks. Moreover
since the string endpoints are fixed at a finite radial distance it can be checked at the
probe string does not reach the gauge theory brane along a normal direction.
\begin{figure}
\begin{minipage}{7cm} 
\bc
\includegraphics[width=7.5cm]{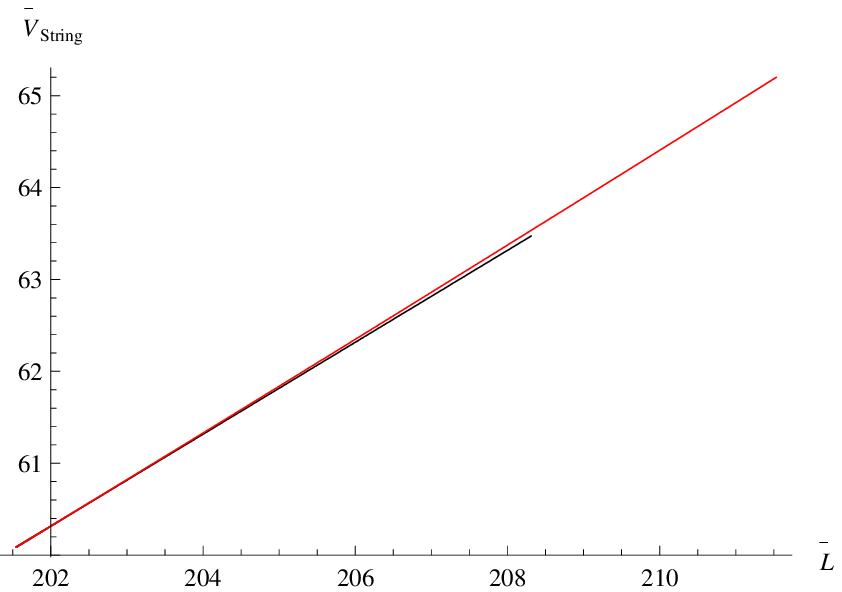}
\caption{Double valued $V_{\sf string}(L)$ relation for gMN background with $\mu=-1$. The
upper red curve is the unphysical branch corresponding to string embeddings to the right
of the minimum in figure \ref{LMNgmumenosuno}. The curve does not reach the origin, manifesting
a minimum quark separation length.}
\label{gancho}
\ec
\end{minipage}\  \
\hfill
\begin{minipage} {7cm} 
\begin{center}
\includegraphics[width=7.5cm]{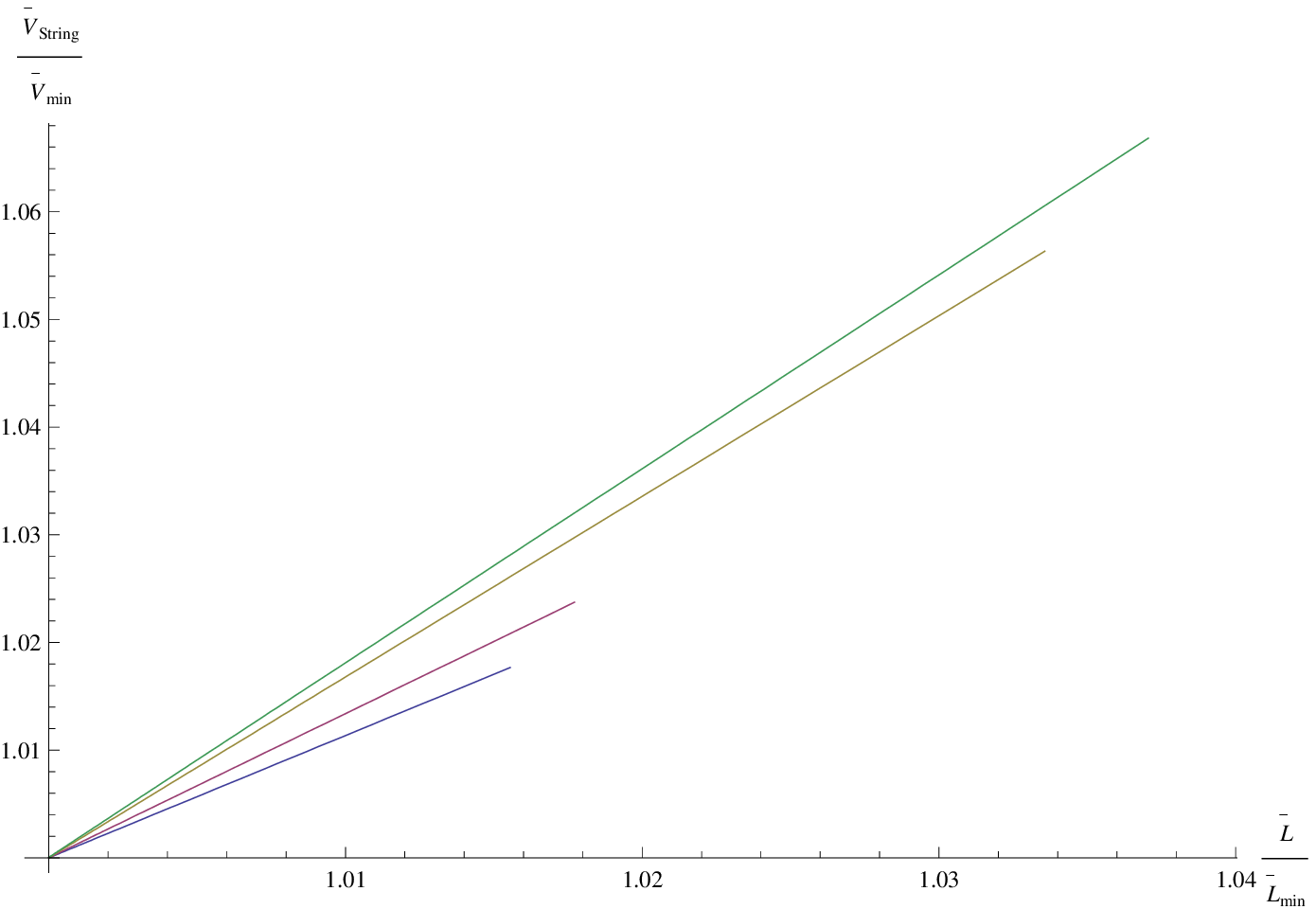}
\caption{Normalized linear $\bar V_{\sf string}(L)$ relation for the physical (left) branches
of figure \ref{LMNgmumenosuno}. Colors as in fig. \ref{LMNgmumenosuno}. }
\label{elvarios}
\end{center}
\end{minipage}
\end{figure}

In figures \ref{LMNgmumenosuno} and \ref{EMNgmumenosuno} we plot  numerical solutions of (\ref{LMNgen}) and (\ref{EMNgen})
for various values of $\mu$. Figure (\ref{LMNgmumenosuno})
shows that $L(\rho_0)$ attains a global minimum for all values of $\mu$ (except $\mu\ne-\frac23$). In other words, no solution exists
for quarks separations $L<L_{min}$.  It is interesting to note
that the minimum value for $L$ is attained, for all values of $\mu$, in a rather small
region in the $\rho$ coordinate near the origin. Based on the concavity considerations
discussed at the end of section \ref{wilson} we expect
the string worldsheets to the left of the minimum $\rho_c$ to be physically
meaningful (stable) and the ones to the right of $\rho_c$ to be unphysical (unstable).
We will show in the following section that this is in fact the case by analyzing quadratic
fluctuations around the solutions. Negative eigenvalues appear for the
$L'(\rho_0)>0$ branch of solutions.

We finally  plot in figures (\ref{gancho}) and (\ref{elvarios}) the $V_{\sf string}(L)$ relation for the generalized MN backgrounds
where a linear confining behavior is observed.
Figure (\ref{gancho}) shows the double valued $V_{\sf string}(L)$ relation for $\mu=-1$, the upper red branch
(unphysical) corresponds to the string configurations to the right of the minimum in fig. (\ref{LMNgmumenosuno}).
In figure \ref{elvarios} we show the $V_{\sf string}(L)$ relation for the physical branches of figure
\ref{LMNgmumenosuno} for several values of $\mu$. Proceeding as in (\ref{energMN}) one finds that all
solutions lead to a $\mu$-independent string tension
\be
T_{\sf string}
=\frac {g_s}{2\pi\alpha'}\,.
\ee

\section{Stability Analysis}
\label{stab} In this section we study, for the backgrounds presented in the previous
section, the eigenvalue problem given
by the equation of motion (\ref{SL}) for in-plane fluctuations in
the $r$-gauge.  We are
interested in searching for unstable modes. For the reasons
discussed at the end of section \ref{wilson}, our aim is to show
that negative eigenvalues ($\omega^2<0$) appear for string
configurations belonging to regions where $L'(r_0)>0$. We
study even solutions, this means that we choose $C'_1=1$ and $C'_0=0$ in
(\ref{asymr0}) as the initial condition at the tip\footnote{Even solutions correspond to arbitrary $C'_1$ at the tip,
its value fixes the normalization of the solution.}. We numerically implement
this conditions as (see (\ref{bc}))
\be
     \delta x_1(r)+2(r-r_0)\frac{d\delta x_1(r)}{dr}=0, ~~~~r\rightarrow
     r_0\nn
\ee
\be
~~~~~~~~~~~~~~~~~\sqrt{r-r_0}\,\delta x_1(r)=1\,
,~~~~r\rightarrow r_0\,.
\label{bcs}
\ee
Solving numerically, the allowed eigenvalues $\omega^2$ for (\ref{SL}) are obtained by demanding $\delta x_1(r)$
to be a normalizable solution
\be
\delta x_1(r)= 0\,, ~~~~~~r\to\infty.
\ee
For completeness we recall now the relation between zero modes for in-plane modes and critical points of the
$L(r_0)$ function \cite{avramis}. The zero mode solution of (\ref{SL}) can be immediately written down
\be
\delta x_1^{(0)}(r)=C\int_r^\infty d\bar r\frac{g(\bar r)f(\bar r)}{(f^2(\bar r)-f^2(r_0))^{\frac32}}+C'
\label{zero}
\ee
where $C',C$ are integration constants, $C'=0$ to get  a normalizable solution and we set $C'=1$. Integrating by parts
in (\ref{zero}) and using (\ref{Lprima}) one obtains
\bea
\delta x_1^{(0)}(r)&=&-\int_r^\infty d\bar r\frac{g(\bar r)}{f'(\bar r)}\frac{d}{d\bar r}\left(\frac1{\sqrt{f^2(\bar r)-f^2(r_0)}}\right)\nn\\
&=&\frac{g(r)}{f'(r)\sqrt{f^2(r)-f^2(r_0)}}+\frac{L'(r)}{2f'(r)} \,.
\eea
Expanding this last expression around the tip $r=r_0$ one has
\be
\delta x_1^{(0)}(r)=\frac{g(r_0)}{\sqrt2(f'(r_0))^{\frac32}}\frac1{\sqrt{r-r_0}}+\frac{L'(r_0)}{2f'(r_0)}+{\cal O}\left(\sqrt{r-r_0}\right)\,.
\label{expzero}
\ee
Generically the first factor in the rhs of (\ref{expzero}) is non-zero, so a necessary and sufficient  condition for
obtaining an even zero mode solution (see (\ref{bcs})) requires the second term in (\ref{expzero}) to cancel,
equivalently $r_0$ must be  a critical point of the $L(r_0)$ length function \cite{avramis}.

\subsection{$AdS_5$}

\begin{figure}
\begin{minipage}{7cm} 
\bc
\includegraphics[width=7.5cm]{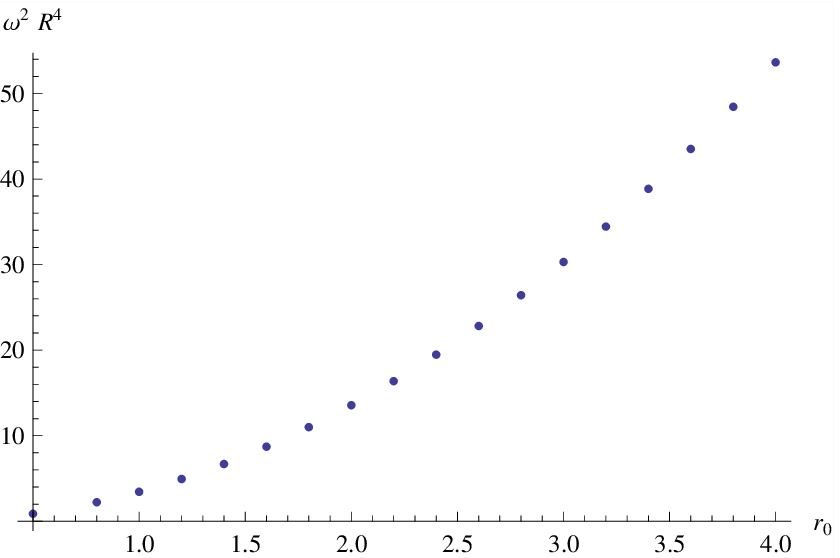}
\caption{Lowest numerical eigenvalue $\omega^2$ of (\ref{fluctads5})
giving a normalizable solution as a function of $r_0$.}
\label{wvsrminAdS}
\ec
\end{minipage}\  \
\hfill
\begin{minipage}{7cm} 
\begin{center}
\includegraphics[width=7.5cm]{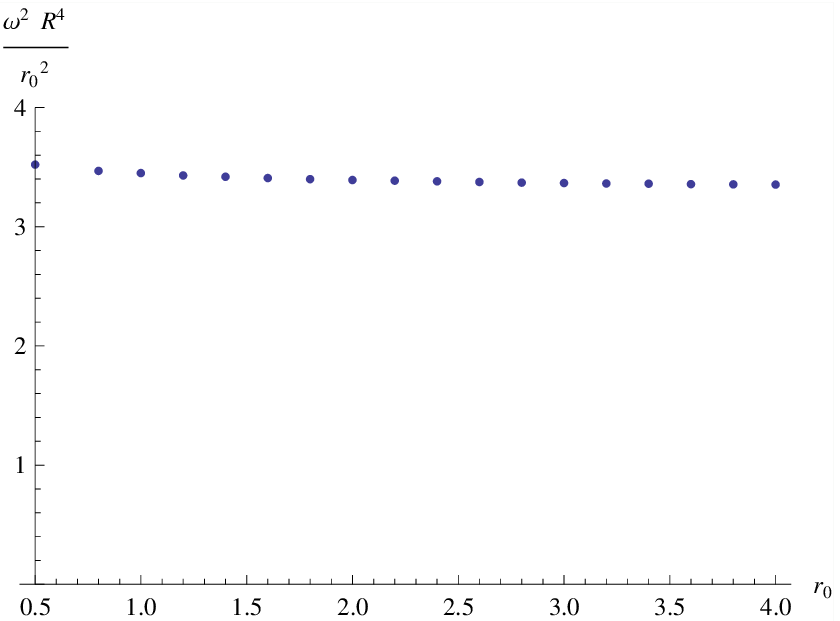}
\caption{Properly normalized lowest eigenvalues of fig. \ref{wvsrminAdS}  as
a function of $r_0$ (see (\ref{fluctnormalized})).}
\label{wvsrminAdS2}
\end{center}
\end{minipage}
\end{figure}
The in-plane fluctuations equation of motion (\ref{SL}) for the $AdS$
spacetime in Poincare coordinates (\ref{ads5}) takes the form
\be
\left[\frac{d}{dr}\left(\frac{(r^4-r_0^4)^{\frac32}}{r^2}\frac{d}{dr}\right)+
\omega^2R^4\frac{\sqrt{r^4-r_0^4}}{r^2}\right]\delta x_1(r)=0\,~~~~0<r_0\le r<\infty\,.
\label{fluctads5}
\ee
Dilatation invariance implies that one should be able to factor out the $r_0$ dependence.
Making $r=r_0\, \rho$ one obtains
\be
\left[\frac{d}{d\rho}\left(\frac{(\rho^4-1)^{\frac32}}{\rho^2}\frac{d}{d\rho}\right)+
\frac{\omega^2R^4}{r_0^2}\frac{\sqrt{\rho^4-1}}{\rho^2}\right]\delta x_1(\rho)=0\,.
\label{fluctnormalized}
\ee
The asymptotic behavior ($\rho\rightarrow\infty$) of (\ref{fluctnormalized}) reads
\be
\left[\frac{d}{d\rho}\left(\rho^4\frac{d}{d\rho}\right)+\frac{\omega^2R^4}{r_0^2}\right]\delta x_1(\rho)\approx0\,,
~~~~~~\rho\gg 1\,,
\label{asymAds}
\ee
whose solutions are
\be
\delta x_1(\rho)\approx\alpha_0+\frac{\alpha_1}{\rho^3}\,,~~~~~~\rho\gg 1\,,
\label{asymads2}
\ee
with $\alpha_0,\,\alpha_1$ integration constants. The  behavior (\ref{asymads2})
implies that normalizable solutions ($\alpha_0=0$) will be found only for particular (discrete)
values $\omega_n^2$.

As a test for our shooting method, we have numerically integrated (\ref{fluctads5}) for
different values of $r_0$ and determined the minimum $\omega^2$ eigenvalues
leading to a normalizable solution. In figure \ref{wvsrminAdS}
we plot these $\omega^2$ as a function of $r_0$. They are
positive for all values of $r_0$, signaling the stability of the U-shaped string
configuration. In fig. \ref{wvsrminAdS2} we show the expected $r_0$-independence of
the mode when properly normalized (see (\ref{fluctnormalized})).
In the following table we show the first eigenvalues corresponding to
even boundary conditions at the tip.
\bc
\begin{tabular}{|r||c|l|}
 \hline
    &   ${\omega_n^2R^4}/{r_0^2}$\\
     \hline
  $n=1$  & ~3.450\\
  $n=3$  & ~22.113\\
  $n=5$  & ~52.325\\
  $n=7$  & ~94.558\\
  $n=9$  & ~148.845\\  \hline
\end{tabular}
\ec
In section \ref{scho} we prove the stability of the configuration by transforming the differential
equation (\ref{fluctads5}) into a  Schrodinger like one (see appendix \ref{sl2sc}).

\subsection{Non-Extremal $D3$-branes}
\label{nond3}

\begin{figure}[h]
\bc
\includegraphics[width=7.5cm]{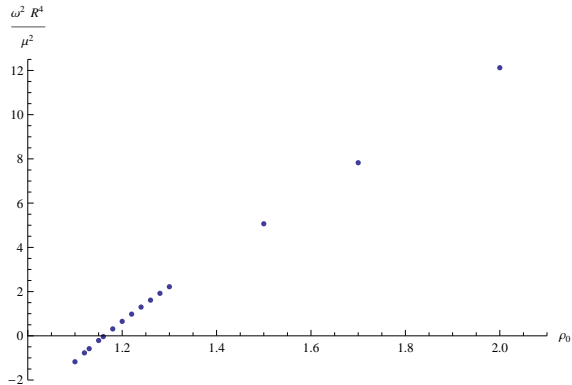}
\caption{Lowest $\omega^2$ eigenvalue of (\ref{eqdiff}) giving a normalizable solution as a function of $\rho_0$. A zero mode
appears for  $\rho_0\simeq1.177$. The classical solutions with $\rho_0<1.177$ are unstable against linear perturbations.}
\label{wvsrminnonextremal}
\ec
\end{figure}

The in-plane fluctuation equation of motion (\ref{SL}) for the background (\ref{thermalads}) takes the form
\be
\left[\frac{d}{d\rho}\left(\frac{(\rho^4-\rho_0^4)^{\frac32}}{\sqrt{\rho^4-1}}\frac{d}{d\rho}\right)+
\frac{\omega^2R^4}{\mu^2}\frac{\rho^4\sqrt{\rho^4-\rho_0^4}}{(\rho^4-1)^{\frac32}}\right]\delta x_1(\rho)=0\,,~~~~1<\rho_0\le \rho<\infty\,.
\label{eqdiff}
\ee
The background (\ref{thermalads}) asymptotes $AdS$
and  therefore the asymptotic  ($\rho\rightarrow\infty$) behavior of the solutions
of (\ref{eqdiff}) is given by (\ref{asymads2}). As in the case of the last subsection,
we expect to find a discrete set of eigenvalues leading to normalizable solutions.

We plot in figure \ref{wvsrminnonextremal} the lowest eigenvalue we found when numerically solving (\ref{eqdiff}) looking for normalizabe
solutions. A zero mode appears precisely at the critical point of the length function $L(\rho_0)$,
that is for $\rho_0\simeq1.177$ (see fig. \ref{L(r_0)}),
in agreement with (\ref{expzero}).

We conclude that the (left) branch in figure \ref{L(r_0)} having $L'(\rho_0)>0$ is unstable under linear perturbations. Finally, note that
the numerical analysis indicates that solutions on the (right) branch in figure \ref{L(r_0)} having $L'(\rho_0)<0$ are stable under linear
perturbations. Nevertheless, as discussed at the end of section \ref{adssch} one expects the solutions with $1.177<\rho_0<1.524$ to be
metastable, decaying to a pair of free quarks.

\subsection{Maldacena-N\'u{\~n}ez background}
\label{mnbkg}

The in-plane fluctuation equation of motion for the Maldacena-N\'u\~nez background
(\ref{metric}) takes the following form (\ref{SL})
\be
\left[\frac{d}{dr}\left(\frac{(e^{2\phi(r)}-e^{2\phi(r_0)})^{\frac32}}{e^{2\phi(r)}}
\frac{d}{dr}\right)+\bar \omega^2\sqrt{e^{2\phi(r)}-e^{2\phi(r_0)}}\right]\delta
x_1(r)=0\,,~~~~0<r_0\le r<\infty\,.
\label{SLMN}
\ee
where $\bar \omega^2=\omega^2\alpha' N$. We now compute the asymptotic behavior of (\ref{SLMN}) to see whether
we should expect a
quantized spectrum or not. In the $r\rightarrow \infty$ limit the equation
of motion for $\delta x_1(r)$ reads
\be
\left[\frac{d}{dr}\left(e^rr^{-\frac14}\frac{d}{dr}\right)+\bar\omega^2e^rr^{-\frac14}\right]\delta
x_1(r)=0\,,~~~~~~r\gg 1\,,
\label{asymMNinfinity}
\ee
where we used that $e^{2\phi(r)}\rightarrow e^{2r}r^{-\frac12}$ for $r\gg1$. This last equation can be
written as
$$
\left[\frac{d^2}{dr^2}+\left(1-\frac{1}{4r}\right)\frac{d}{dr}+\bar \omega^2\right]\delta x_1(r)=0\,,~~~~~~r\gg 1\,.
$$
The $r^{-1}$ term can be omitted in the large $r$ limit and the
asymptotic solution to (\ref{SLMN}) is then
\be
\delta x_1(r)\simeq e^{-\frac12r}(\beta_0\, e^{r\alpha}+\beta_1\,e^{-r\alpha})\,,~~~~~~r\gg 1\,,
\label{asymMN}
\ee
where $\alpha=\frac{\sqrt{1-4\bar\omega^2}}{2}$. From (\ref{asymMN})
it follows that any $\bar \omega^2>0$ lead to normalizable solutions, the spectrum of stable
in-plane fluctuation is therefore continuum.
In the $\bar\omega^2\leq0$ case ($\alpha\ge\frac12$) $\beta_0$ must be set to zero and we have the
possibility of getting a discrete spectrum of negative eigenvalues. Our
numerical analysis could not find any normalizable negative eigenmodes, suggesting the stability of the
classical configuration in agreement with the concavity condition (\ref{convexity}).

In the section \ref{scho} we show that negative eigenvalues does not exist from the study of a Schrodinger
equation analysis of (\ref{SLMN}) (see appendix \ref{sl2sc}).

\subsection{Klebanov-Strassler background}
\label{ksbkgd}

The equation of motion for the in-plane fluctuation in this case  takes the form
\be
\left[\frac{d}{dr}\left(\frac{K(r)}{h(r)}\left(1-\frac{h(r)}{h(r_0)}\right)^{\frac32}\frac{d}{dr}\right)+
\bar \omega^2\frac1{6K(r)}\sqrt{1-\frac{h(r)}{h(r_0)}}\right]\delta x_1(r)=0\,,~~~~0<r_0\le r<\infty\,,
\label{slks}
\ee
with $\bar \omega^2$ dimensionless and $K(r)$ and $h(r)$ given by (\ref{kks}) and (\ref{hks}) respectively.
From the $r\rightarrow\infty$ limit of $K(r)$ and $h(r)$  one obtains
\be
\left[\frac{d}{dr}\left(\frac{e^{r}}{r}\frac{d}{dr}\right)+\bar\omega^2\frac{e^{\frac{r}{3}}}{2^{\frac43}}\right]\delta x_1(r)=0\,,~~~~r\gg 1\,,
\ee
which gives
\be
\left[\frac{d^2}{dr^2}+\left(1-\frac1r\right) \frac{d}{dr}+\bar \omega^2 \frac{re^{-\frac23r}}{2^{\frac43}}\right]\delta x_1(r)=0\,,~~~~r\gg 1\,.
\label{asymKS}
\ee
In the large $r$ limit the $r^{-1}$ and the last term in (\ref{asymKS}) can be omitted and
the in-plane fluctuation asymptotics turns to be
\be
 \delta x_1(r)\simeq \alpha_0+\alpha_1e^{-r}\,,~~~~~~r\gg 1\,.
\ee
The integration constant $\alpha_0$ must be set to zero to obtain normalizable solutions and we therefore expect
to get a discrete eigenvalue spectrum.
\begin{figure}[h]
\bc
\includegraphics[width=7.5cm]{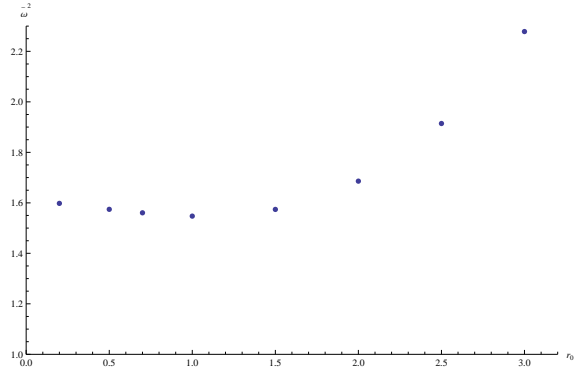}
\caption{Numerical solution for the lowest $\omega^2$ leading to a
normalizable solution as a function of $r_0$ for the Klebanov-Strassler background. No
negative eigenvalues where found.}
\label{wvsrminKS}
\ec
\end{figure}
In figure (\ref{wvsrminKS}) we plot the lowest eigenvalue of (\ref{slks}) we have found numerically leading to a normalizable
solution as a
function of $r_0$. We have not found numerically any negative eigenvalues. In section \ref{scho}  we will prove
the stability of the classical solution transforming (\ref{slks}) to a
Schrodinger like equation and showing that no negative modes can appear (see appendix \ref{sl2sc}).

\subsection{Generalized Maldacena-N\'u\~nez}
\label{gMNbkgd}

The equation of motion for the in-plane fluctuation $\delta x_1(\rho)$ in the backgrounds (\ref{gMN}) takes the form
\be
\left[\frac{d}{d\rho}\left(\frac{(e^{8f(\rho)}-e^{8f(\rho_0)})^{\frac32}}{e^{8f(\rho)+k(\rho)}}\frac{d}{d\rho}\right)+
\bar\omega^2e^{k(\rho)}\sqrt{e^{8f(\rho)}-e^{8f(\rho_0)}}\right]\delta x_1(\rho)=0\,,~~~~0<\rho_0\le \rho<\infty\,.
\label{SLMNgen}
\ee
In the  $\mu=-\frac23$ case ($k(\rho)=const.$) the equation (\ref{SLMN}) for the Maldacena-Nu{\~n}ez background  is recovered
(from now on we consider $\mu\ne-\frac{2}{3}$).
In the large $\rho$ limit the gMN solutions asymptote the deformed conifold and the $f$ function approaches a constant $f_\infty$, the
asymptotic behavior is then given by
\be
\left[e^{-k(\rho)}\frac{d}{d\rho}\left(e^{-k(\rho)}\frac{d}{d\rho}\right)+\bar\omega^2\frac{e^{8f_\infty}}
{e^{8f_\infty}-e^{8f(\rho_0)}}\right]\delta x_1(\rho)=0\,,~~~~\rho\gg 1\,.
\label{asymgMN}
\ee
Returning to the original $r$ variable in (\ref{gMN}) ($dr=e^{k(\rho)}d\rho$) one obtains
\be
\left[\frac{d^2}{dr^2}+\tilde\omega^2\right]\delta x_1(r)=0\,,~~~~r\gg 1\,.
\ee
whose solutions are plane waves $e^{\pm i\tilde\omega r}$ for $\bar\omega^2>0$ and real exponentials $e^{\pm\tilde\omega r}$ for $\bar\omega^2<0$ case.
We conclude that no normalizable solutions exist for $\bar\omega^2>0$. A word of caution, as discussed in section \ref{gmnsol} the gauge theory
brane must be placed at a finite distance $\rho_\infty$, therefore  for (\ref{SLMNgen}) defined on
$\rho_0\le\rho\le \rho_\infty$ positive eigenvalues will exist.
In the $\bar\omega^2<0$ case the possibility for
negative eigenmodes exists and in fact we find normalizable negative mode solutions precisely for the classical solutions region where the convexity
condition (\ref{convexity}) is not satisfied.
\begin{figure}
\begin{minipage}{7cm} 
\bc
\includegraphics[width=7.5cm]{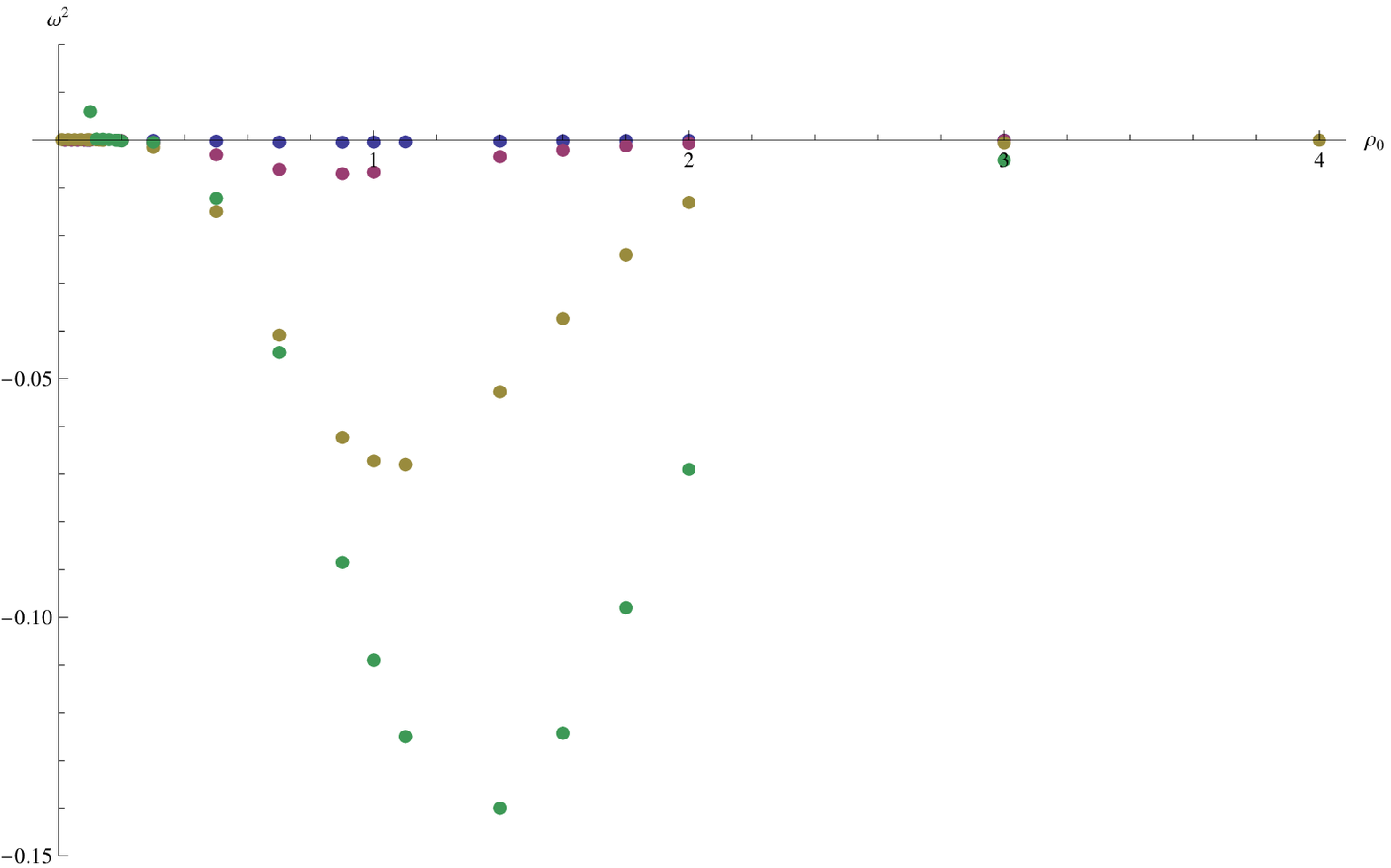}
\caption{Lowest $\omega^2$ eigenvalue of (\ref{SLMNgen}) leading to a normalizable solution as a function of $\rho_0$.
Negative (unstable) modes are found precisely for the classical embeddings satisfying $L'(\rho_0)>0$. Colors as in fig. \ref{LMNgmumenosuno}.}
\label{wvsrminMNgenmu}
\ec
\end{minipage}\  \
\hfill
\begin{minipage}{7cm} 
\begin{center}
\includegraphics[width=7.5cm]{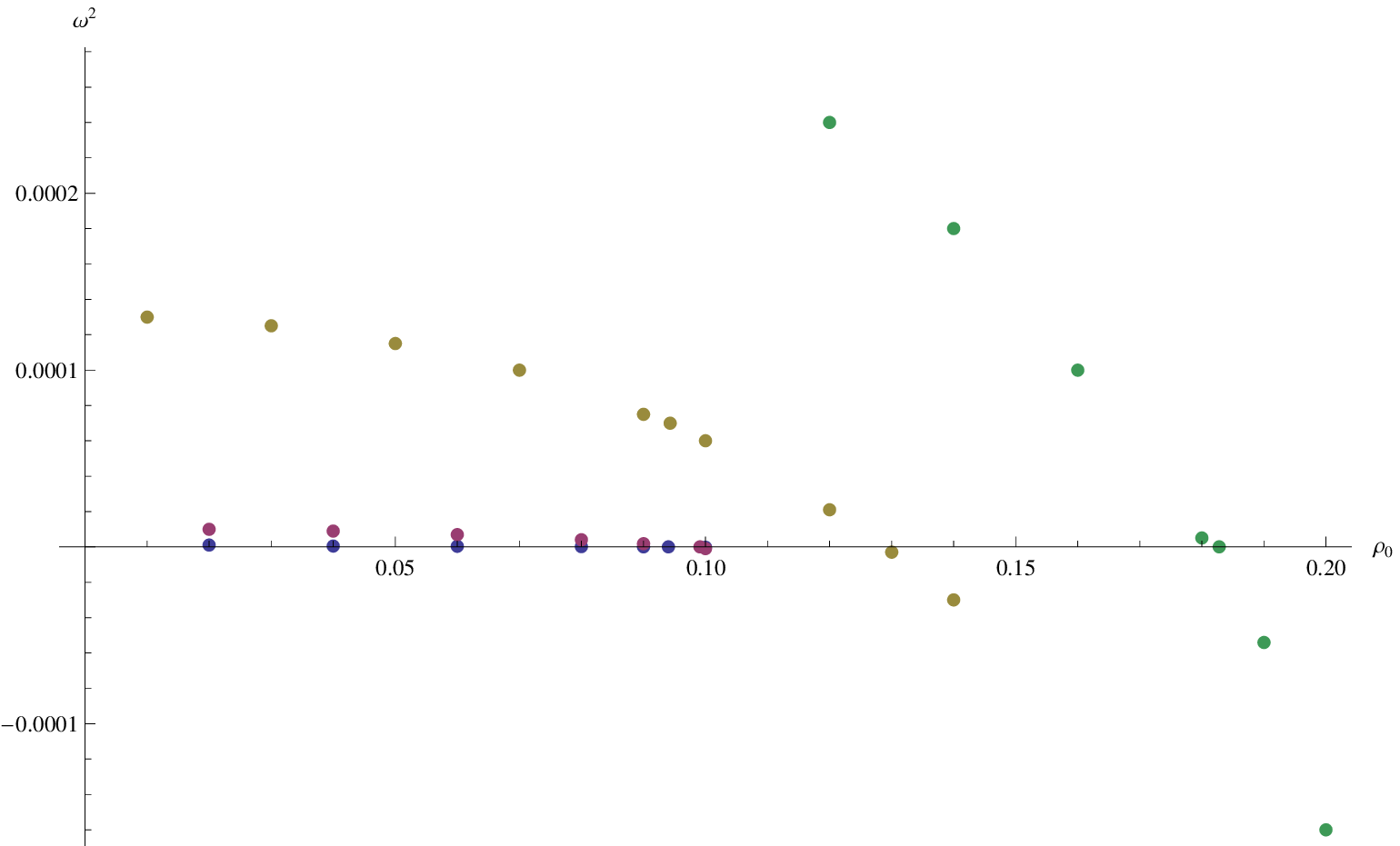}
\caption{Zoom of figure \ref{wvsrminMNgenmu} near the origin. The lowest eigenvalues are positive for $L'(\rho_0)<0$ solutions. A zero mode
appears precisely for the critical values of the length function $L(\rho_0)$ (see fig. \ref{LMNgmumenosuno}).}
\label{wvsrminMNgenzoommu}
\end{center}
\end{minipage}
\end{figure}
In the figures \ref{wvsrminMNgenmu} and \ref{wvsrminMNgenzoommu} we plot the minimal eigenvalues leading to normalizable solutions
we found numerically as a function of $r_0$. We found complete agreement with figure \ref{LMNgmumenosuno}: no instabilities are
found for classical solutions satisfying $L'(r_0)<0$, on the other hand, we find negative (unstable) modes for the right branch curve
($L'(r_0)>0$ solutions) in figure \ref{LMNgmumenosuno}. These
results are gratifying since unstable modes are found precisely for the classical embeddings which do not satisfy the conditions (\ref{convexity}).
In the following section we review this results by transforming the equation into a Schrodinger like problem.

\section{Schrodinger Potentials Analysis}
\label{scho}

In this section we analyze the fluctuation equation of motion (\ref{SL}) transforming it to
a Schrodinger like equation (see appendix \ref{sl2sc}). From the form of the potential it is possible
in some cases to show that no negative eigenvalues can appear and therefore to prove the
stability of the corresponding classical embeddings.

\subsection{$AdS_5$}

\begin{figure}
\begin{minipage}{7cm} 
\bc
\includegraphics[width=7.5cm]{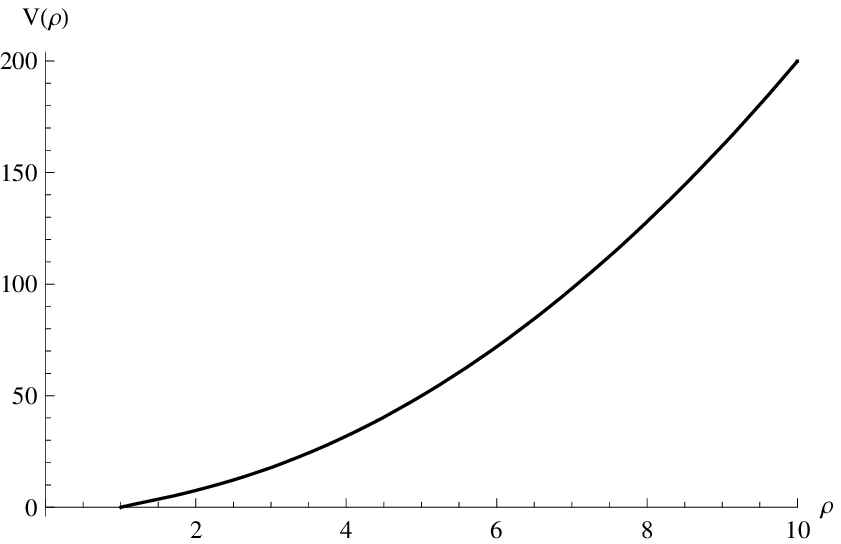}
\caption{Schrodinger potential (\ref{potads}) for
in-plane fluctuations in $AdS$ the positive definite property of it
guarantees that no negative eigenmodes exist.}
\label{potschrads5s5}
\ec
\end{minipage}\  \
\hfill
\begin{minipage}{7cm} 
\begin{center}
\includegraphics[width=7.5cm]{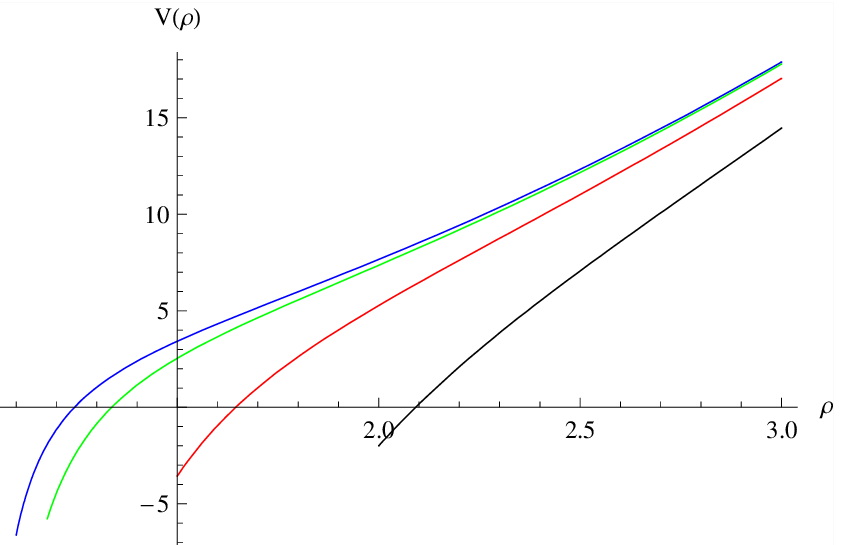}
\caption{$V(\rho,\rho_0)$ (eqn.(\ref{pottads})) for different  $\rho_0$ va\-lues.
The blue  line corresponds to $\rho_0=1.1$, the green to the critical value
$\rho_0=1.177$ and  the red, black lines to $\rho_0=1,\,2$. The region where the potential
is negative diminishes as $\rho_0$  increases. Negative eigenmodes cease to exist for $\rho_0\ge1.177$}
\label{potschrnonextremal}
\end{center}
\end{minipage}
\end{figure}

The Schrodinger potential (\ref{schrodinger}) for the equation (\ref{fluctnormalized}) takes form \cite{avramis}
\be
V(\rho)=2\,\frac{\rho^4-1}{\rho^2}\,,~~~~\rho\in[1,\infty)\,,
\label{potads}
\ee
here $\rho$ should be understood as $\rho=\rho(y)$. The change of variables (\ref{chv}) leading to the Schrodinger equation
(\ref{sch}) can be analytically computed
\be
y(\rho)=y_0-\frac14\, \mathsf{B}\left(\frac1{\rho^4};\frac14,\frac12\right)\,,
\label{chvads}
\ee
with $y_0=\frac{\Gamma[\frac14]^2}{4\sqrt{2\pi}}$. The half line $\rho\in[1,\infty)$ of the original Sturm-Liouville
problem, under the change of variables (\ref{chvads}), maps to the finite interval $y\in[0,y_0]$, the potential (\ref{potads}) diverging
at $y_0$. We have therefore obtained a Schrodinger problem defined on a finite interval with canonical boundary
conditions (see (\ref{bcsch})-(\ref{bcsschrod})) hence a discrete spectrum
will result, moreover, since the potential (\ref{potads}) is positive definite a standard QM argument tell us that
no negative eigenvalue solutions exist. We conclude that the $AdS$ embedding given by (\ref{xads}) is
stable under linear perturbations. Figure \ref{potschrads5s5} shows a plot of the potential (\ref{potads}) as a function
of $\rho$, the true variable for the Schrodinger problem is $y$ given by (\ref{chvads}) and it amounts to a rescaling
of the horizontal axis in fig. \ref{potschrads5s5} mapping $\rho=\infty$ to a finite distance.

\subsection{$AdS_5$-Schwarzschild}

The Schrodinger potential (\ref{schrodinger}) for the in-plane fluctuation equation (\ref{eqdiff})
takes form \cite{avramis}
\be
V(\rho,\rho_0)=2\,\frac{\rho^8(\rho^{4}-\rho_0^{4})-\rho_0^4(4\rho^4-1)-3\rho^4}{\rho^6(\rho^4-1)}\,,~~~~1<\rho_0\le \rho<\infty\,.
\label{pottads}
\ee
The behavior of this potential for different values of $r_0$ is shown in
figure \ref{potschrnonextremal}. Unlike the $AdS_5\times S^5$ case, there exist regions
where the potential becomes negative, this is in agreement with the results of section \ref{nond3} where
negative eigenmodes where found.
The potential starts from a negative value at $\rho_0$ given by $V(\rho_0,\rho_0)=-8/\rho_0^2$.
As $\rho_0$ increases the negative region gets dimmer and the negative modes cease to exist at the critical
value,  found numerically in section \ref{nond3},  $\rho_{0c}\simeq1.177$ which precisely coincides with
the critical value of the length function $L(\rho_0)$. We conclude that the  classical embeddings satisfying $L'(\rho_0)>0$
are unstable under linear perturbations (see also \cite{avramis} for a perturbative analysis of the eigenvalues). The classical unstable
solutions $L'(\rho_0)>0$ have regularized energy $E_{q\bar{q}}$ greater than zero (see fig. \ref{e(r_0)}), since the reference
configuration satisfies the same boundary conditions, the natural candidate for the decay process is the reference (free quarks) state.

For completeness
we quote that since the asymptotics of this background coincides with the previous case,  the Schrodinger equation for
in-plane fluctuations results
defined on a finite interval. The spectrum of stable fluctuations is therefore discrete.

\subsection{Maldacena-N\'u\~nez }
\label{schMN}

\begin{figure}
\begin{minipage}{7cm} 
\bc
\includegraphics[width=7.5cm]{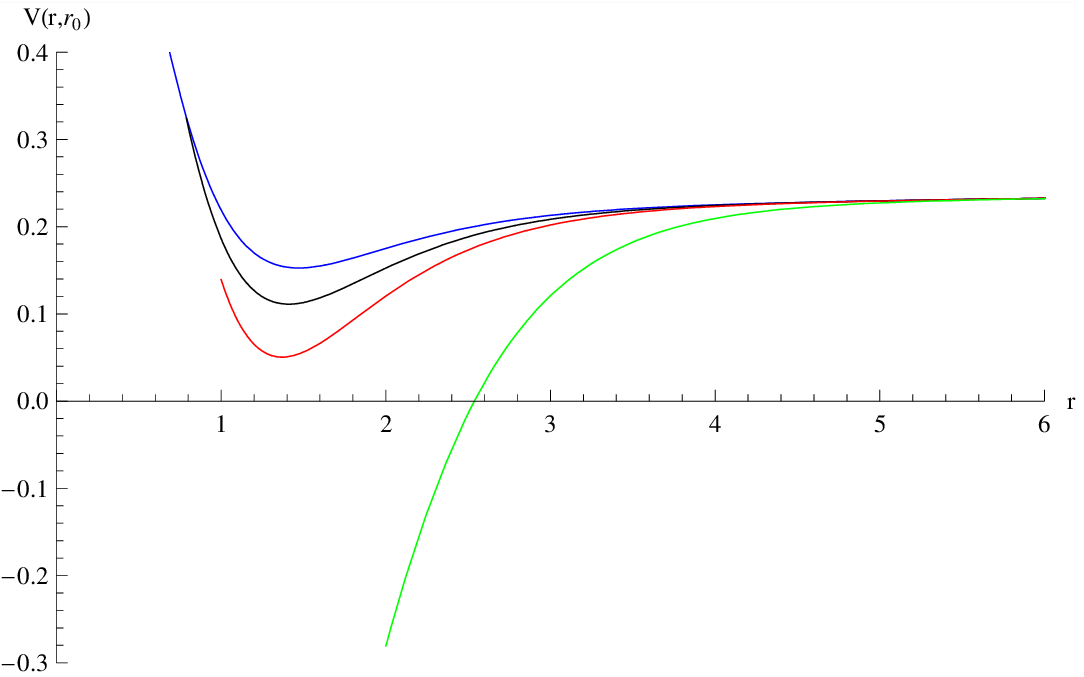}
\caption{Schrodinger potential (\ref{potMN}). The blue, black, red and green lines
corresponds to $r_0=0.2,\,0.7,\,1,\,2$.
The minimum of the potential becomes negative for $r_0\ge1.1605$.}
\label{potschrMN}
\ec
\end{minipage}\  \
\hfill
\begin{minipage}{7cm} 
\begin{center}
\includegraphics[width=7.5cm]{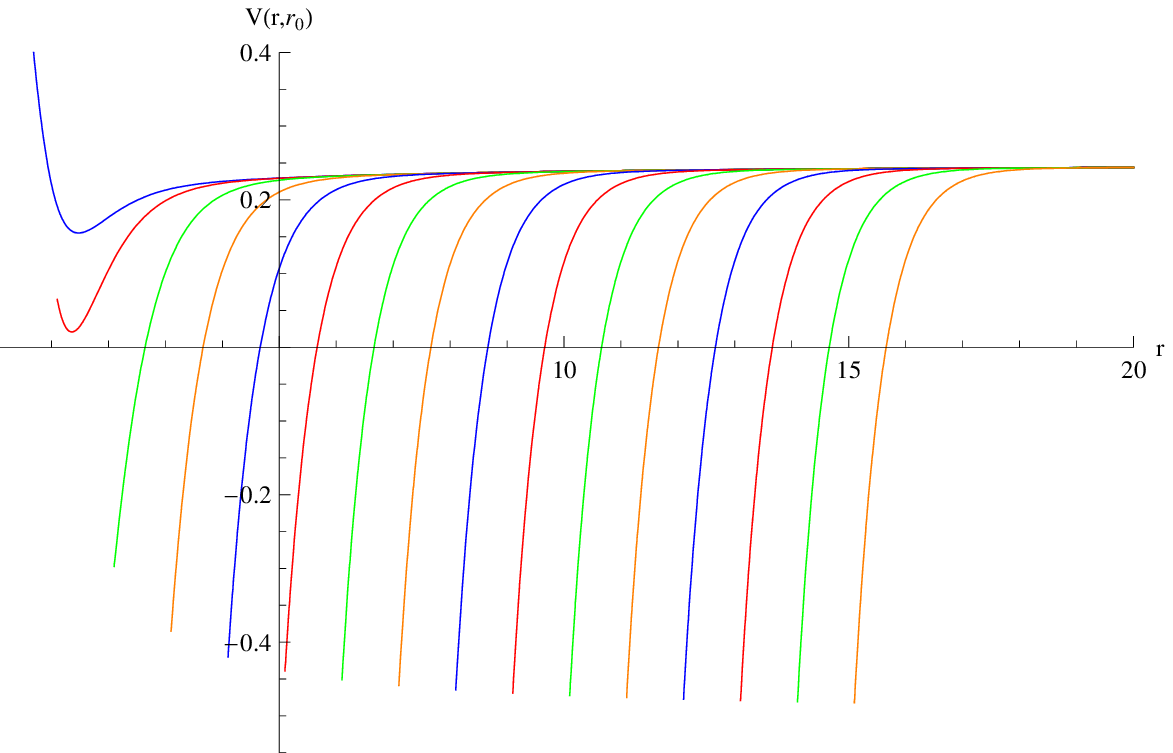}
\caption{Potential (\ref{potMN}) as a function of  $r_0$. The potential asymptotically
tends to $V_\infty=\frac14$ in agreement with (\ref{asymMN}). For $r_0<1.1605$ is positive negative and
 for $r_0\ge 1.1605$ contains
negative regions. Near the tip the potential can be approximated asymptotically (for large $r_0$) by
$V(r)\simeq-\frac12+\frac32 (r-r_0)$}
\label{potenMN}
\end{center}
\end{minipage}
\end{figure}

The Schrodinger potential for (\ref{SLMN}) takes the form
\be
V(r,r_0)=\frac{e^{-2\phi(r)}}{4}\left((e^{2\phi(r)}-3e^{2\phi(r_0)})\phi'^2(r)
+2(e^{2\phi(r)}+e^{2\phi(r_0)})\phi''(r)\right),~~~~0<r_0\le r<\infty\,.
\label{potMN}
\ee
As before $r$ should be understood as $r=r(y)$ and contrary to the last two cases the change of variables (\ref{chv})
give a Schrodinger problem in $y$ coordinates defined on the half line $y\in[0,\infty)$.
Figures \ref{potschrMN} and \ref{potenMN} show the Schrodinger potential (\ref{potMN}) for different values of $r_0$.
We should confront these figures with the results in section \ref{mnbkg}. The outcome of that section,
for all values of $r_0$, was that a continuum
spectrum results for $\omega^2>0$ and numerically no negative normalizable modes were found. We first address the continuum
spectrum for $\omega^2>0$. Figure \ref{potschrMN} shows that the potential is positive definite for $r_0\le1.1605$ and asymptotes
the value $V_{\infty}=\frac14$. One might therefore conclude no solutions for
$0<\bar\omega^2<V_{min}$, a discrete spectrum for $V_{min}<\bar \omega^2<\frac14$ (if possible) and a
continuum, but not normalizable, spectrum for $\bar\omega^2>\frac14$, all in contradiction with
the mentioned results. The agreement is achieved when taking into account the factor $(PQ)^{-\frac14}$
that relates the solution of the Schrodinger equation $\Psi$ with the fluctuation $\delta x_1$ (see appendix \ref{sl2sc} eqn. (\ref{chv}))
\be
\delta x_1=\frac{e^{\frac{\phi(r)}2}}{(e^{2\phi(r)}-e^{2\phi(r_0)})^{\frac12}}\Psi\simeq e^{-\frac{r}2}\Psi\,,~~~~r\to\infty\,.
\ee
The $e^{-\frac{r}2}$ factor makes all  $\bar\omega^2>0$ solutions of the
Schrodinger problem satisfy the $\delta x_1|_{r=\infty}=0$ whether or not they
normalizable as $\Psi(y)$ solutions (asymptotically one has $y\simeq r$). However, for $\bar\omega^2<0$ solutions the
factor is not enough for making the (diverging) solutions satisfy the boundary condition.
We conclude that for all $r_0$ a continuum spectrum results for $\bar\omega^2>0$.

The remaining point to be addressed is the possibility of bound states for $-\frac12<\bar\omega^2<0$
in the limit of large $r_0$. As seen from figure \ref{potenMN}, asymptotically, the potential
starts from $V(r_0)\simeq-\frac12$ and  the linear approximation one obtains is $V(r)\simeq-\frac12+\frac32 (r-r_0)$.
The relation between the $r$ and $ y$ coordinates (\ref{chv}) in the same limit is $(r-r_0)\simeq y^2/2$. All
these leads to a harmonic oscillator in $y$ coordinates with bound state energy above zero. We
therefore conclude that no bound states exist. We find a complete
agreement between the Schrodinger analysis and the numerical results of section \ref{mnbkg}.

The results are appealing since if instabilities were found, no obvious candidate for the
decay is available (cf. section \ref{schgMN}).

\subsection{Klebanov-Strassler}
\label{schKS}

\begin{figure}[h]
\hspace{-0.25cm}
\begin{minipage}[b]{1\linewidth} 
\centering
\includegraphics[width=7.5cm]{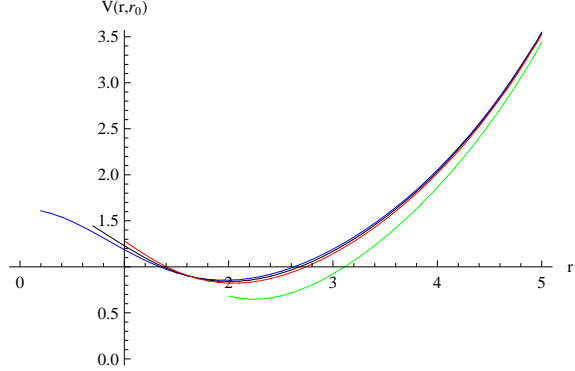}
\caption{Schrodinger potential (\ref{sKS}). The blue, black, red and green lines corresponds
to $r_0=0.2,\,0.7,\,1,\,2$. The potential is
positive definite and therefore no negative (unstable) eigenvalue modes result. }
\label{potschrKS}
\end{minipage}
\end{figure}
The potential for the Klebanov-Strassler in-plane fluctuation (\ref{slks}) reads
\bea
V(r,r_0)&=&-\frac{3K(r)}{8{h^{3}(r)h(r_0)}}\left[{4h(r)(h(r)+h(r_0))h'(r)k'(r)}
\right.\nn\\
&&\left.-k(r)(3h(r)+7h(r_0))h'^2(r)+4h(r)(h(r)+h(r_0))h''(r))\right]
\label{sKS}
\eea
The asymptotic behavior of the $P,Q$ functions in this case lead to a Schrodinger
problem defined on a finite $y$ interval (see appendix \ref{sl2sc}). Figure (\ref{potschrKS}) shows the form of the potential for various
$r_0$ values. The  potential is positive definite and  therefore no (unstable) $\omega^2<0$
solutions exist.  The finite interval on which the Schrodinger problem is defined implies
a discrete set of eigenvalues. We find complete agreement with the results of section \ref{ksbkgd}.

\subsection{Generalized Maldacena-N\'u\~nez}
\label{schgMN}
\begin{figure}
\begin{minipage}{7cm} 
\bc
\includegraphics[width=7.5cm]{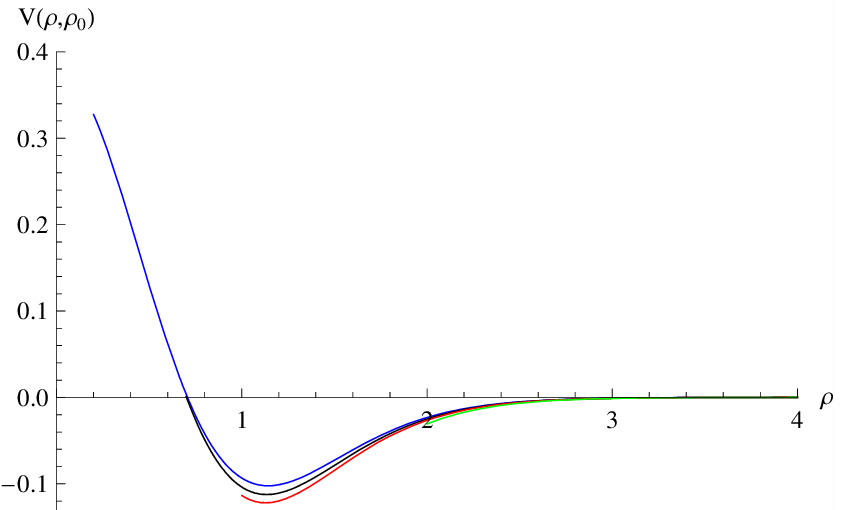}
\caption{Schrodinger potential (\ref{VgMN}) for
different $\rho_0$ values and $\mu=-1$. The blue, black,
red and green lines correspond to $\rho_0=0.2,\,0.7,\,1,\,2$
respectively.}
\label{potschrMNgen}
\ec
\end{minipage}\  \
\hfill
\begin{minipage}{7cm} 
\begin{center}
\includegraphics[width=7.5cm]{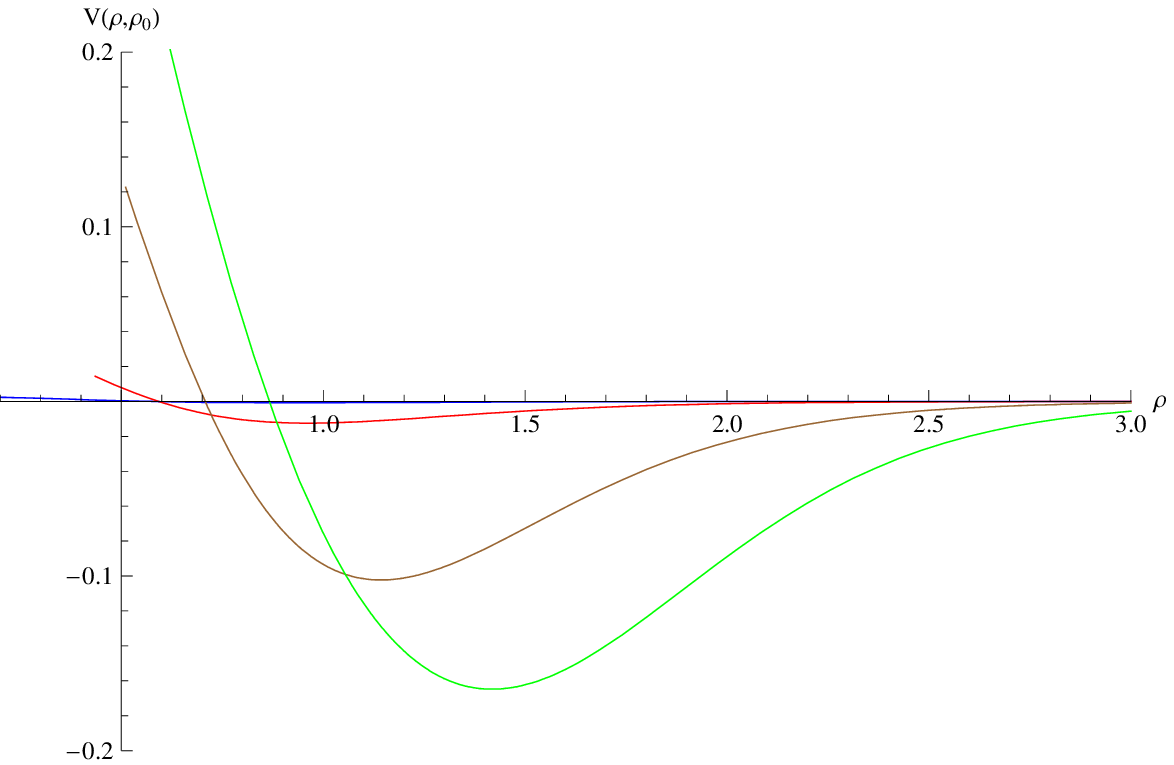}
\caption{Schrodinger potential for $\rho_0=0.2$ and different
values of the parameter $\mu$. The blue, red, brown and green lines
correspond to $\mu=-1.8,\,-1.5,\,-1,\,-0.8$.}
\label{potschrgMNvariosmu}
\end{center}
\end{minipage}
\end{figure}

The Schrodinger potential for in-plane fluctuations in the generalized MN solutions (\ref{SLMNgen}) can be written as
\be
V(\rho,\rho_0)=\frac2{e^{8f(\rho)+2k(\rho)}}\left(2 (e^{8f(\rho)}-e^{8f(\rho_0)})f'^2(\rho)
+(e^{8f(\rho)}+e^{8f(\rho_0)})(f''(\rho)-k'(\rho)f'(\rho))\right)
\label{VgMN}
\ee
The asymptotic behavior of the $P,Q$ functions (see appendix \ref{sl2sc} and eqn. (\ref{asymgMN})) leads to a Schrodinger problem
formulated on the half line $y\in[0,\infty)$. Figure (\ref{potschrMNgen}) shows the  behavior of the potential for different values of $\rho_0$ and
a fixed value of $\mu=-1$. The potential becomes negative above some critical value $\rho_*$ and asymptotes $V_\infty=0$
in concordance  with (\ref{asymgMN}) and  the existence of negative (unstable) modes found numerically in section \ref{gMNbkgd}.
The $(PQ)^{-\frac14}$ factor relating the Schrodinger wave function $\Psi$ to the fluctuation $\delta x_1$
approaches a constant at infinity, therefore not changing the asymptotics of the $\Psi$ solutions (cf sect. \ref{schMN}).
Figure \ref{potschrgMNvariosmu} shows the Schrodinger potential, for a fixed value of $\rho_0=0.2$, for
different values of $\mu$. The minimum  of the potential decreases as the
$\mu$ approaches $-\frac23$. As already mentioned the original MN solution (\ref{metric}) is not continuosly
connected with the generalized class of solutions (\ref{gMN}).
Agreement with the numerical results of section \ref{schgMN} is found but it is no
clear to us which is the final state of the decay.

\section{'t Hooft loop}
\label{thooftloop}

 The electromagnetic dual to  Wilson lines in Yang-Mills theories are the 't Hooft lines \cite{thooftline}.
In four dimensions, the mechanism for confinement is supposed to be due to magnetic monopole condensation
(dual Meissner effect), the analysis in \cite{thooftline} concluded that a screened monopole potential between a
$m\bar m$ pair should be observed when confinement is due to a  dual Meissner effect. A generalization of this idea
is dyon confinement and goes under the name of oblique confinement.

The string prescription for computing 't Hooft loops in the MN and KS solutions was proposed in the same papers \cite{mn}-\cite{ks}
(see \cite{BM} and also \cite{paredes} where a technical issue, correcting the proposed 2-cycle in \cite{mn}, was pointed out) and consists in wrapping
a probe $D3$-brane on the same 2-cycle on which the $D5$-branes leading to
the backreacted geometry were wrapped (see also \cite{groo}). The outcome of the construction is an effective D1-brane (string) which is analyzed
in complete analogy with the probe fundamental string we have been discussing in previous sections. The
important difference with respect to the Wilson loop case is that the 't Hooft loop in generic non S-dual
theories is sensible to the internal five dimensional manifold.

\subsection{Maldacena-N\'u\~nez}

\begin{figure}[h]
\begin{minipage}{7cm}
\begin{center}
\includegraphics[width=7.5cm]{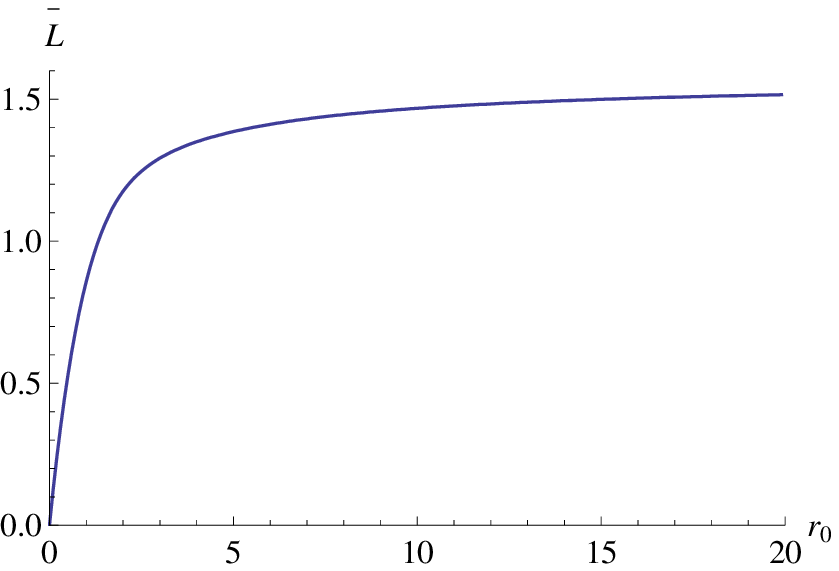}
\caption{The length function for the effective string as a function of $r_0$ in the
Maldacena-N\'u\~nez t Hooft loop case.}
\label{lvsRminMNthooft}
\end{center}
\end{minipage}
\   \
\hfill \begin{minipage}{7cm}
\begin{center}
\includegraphics[width=7.5cm]{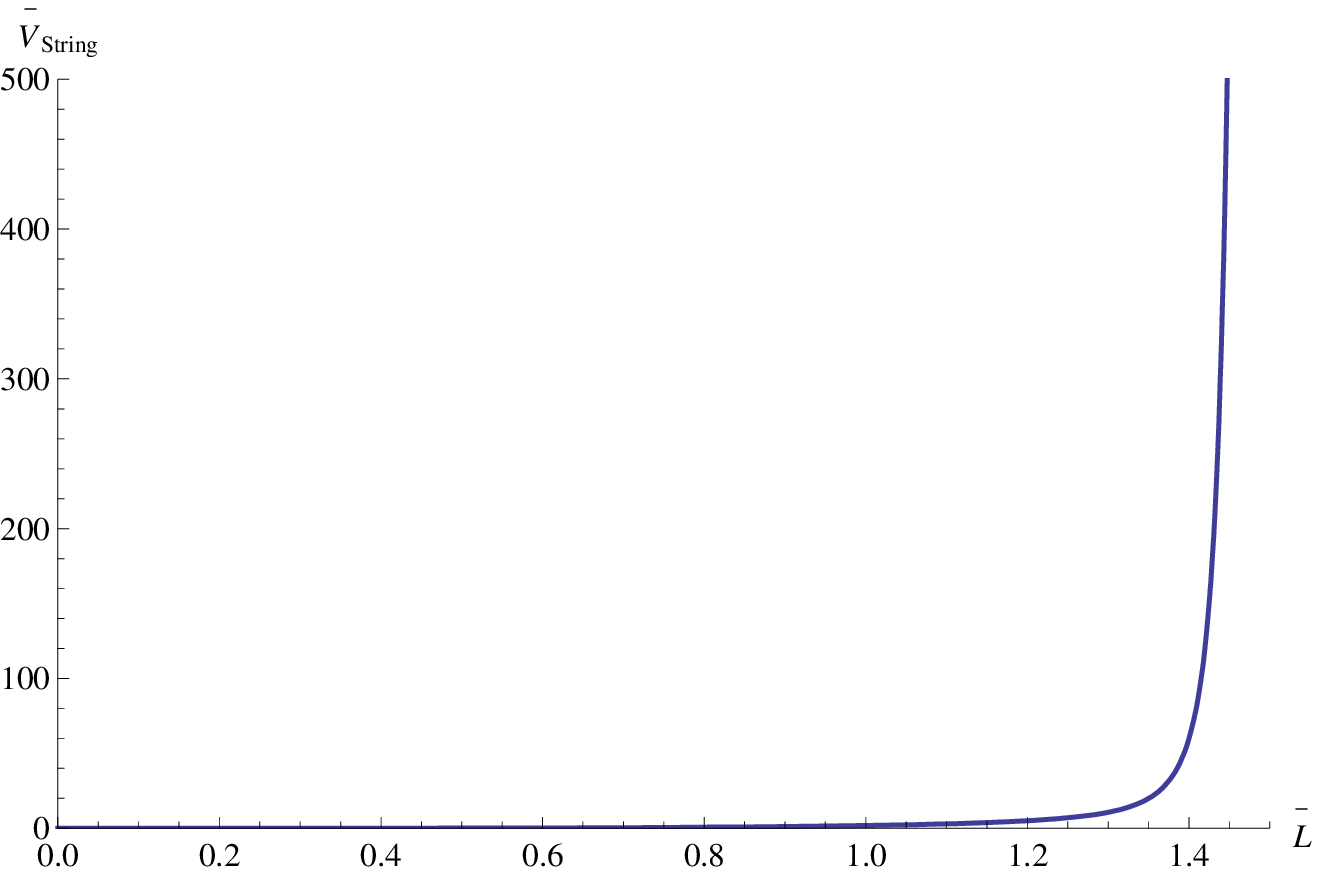}
\caption{ The energy of the monopole-antimonopole pair as a function of the separation length.}
\label{EvsLMNthooft}
\end{center}
\end{minipage}
\end{figure}

The embedding manifold for the $D3$ in the metric (\ref{metric}) is \cite{BM}\footnote{The remaining coordinates are
set to constants. The value of the $\psi$ coordinate is fixed by demanding the $S^2$ to be of minimal volume.}
\be
{\cal M}_4=[t,x,r(x),\theta=\tilde\theta,\varphi=2\pi-\tilde\varphi,\psi=\pi]\,.
\label{ansatz}
\ee
The
induced metric on ${\cal M}_4$ results in
\be
ds^2_{ind}= \alpha'  Ne^{{\phi}}\,\,\Big[-dt^2+(1+\acute r^2)dx^2+(e^{2h}+\frac14(1-a)^2)\,(d\theta^2+\sin^2\theta d\varphi^2)\Big]\,,
\ee
and the expressions (\ref{oneform}) for $a,h$ give
\be
V_{S^2}(r)\equiv \frac14(1-a(r))^2+e^{2h(r)} =r\tanh{r}\,.
\label{s2vol}
\ee
Note that the $S^2$ sphere smoothly collapses at the origin. Integrating  the DBI action\footnote{
Placing
the gauge fixed ansatz (\ref{ansatz}) into the action (\ref{dbi}) give the correct equation of motion
for $r(x)$ which coincides with (\ref{sig}).}
\be
S_{DBI}=-T_{D3}\int d^4\sigma e^{-\phi} \sqrt{g_{ind}}
\label{dbi}
\ee
over the internal manifold ($S^2_{\theta\varphi}$) and the time coordinate one obtains
\be
S_{eff}= 4\pi T_{D3}{\cal T}{ (\alpha'N)^2}\int{ e^\phi\,r\tanh r \sqrt{1+\acute r^2}}\,dx\,.
\ee
The important difference wrt the previous Wilson loop calculation in the
Maldacena-N\'u\~nez background (see sect. \ref{mnsol})  resides in
the  $f(r)$ and $g(r)$ functions in (\ref{lmn})-(\ref{emn}) being multiplied by the
2-sphere volume (\ref{s2vol}).

\begin{figure}[h]
\begin{minipage}{7cm}
\begin{center}
\includegraphics[width=7.5cm]{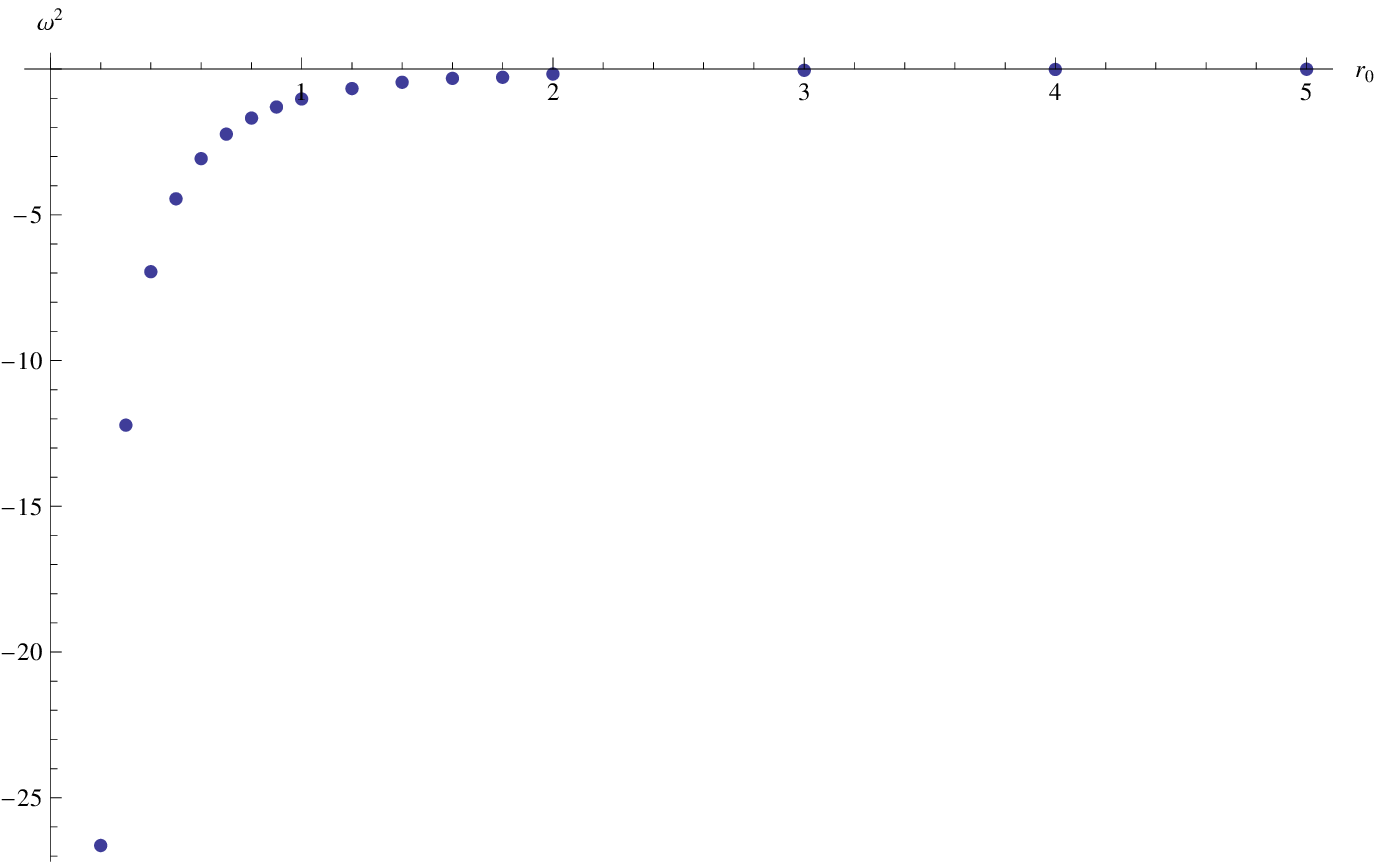}
\caption{Lowest numerical eigenvalue $\omega^2$ for in-plane fluctuations depending only on $t,r$ coordinates
as a function of $r_0$ in the Maldacena-N\'u\~nez 't Hooft loop case.}
\label{wvsrminMNthooft}
\end{center}
\end{minipage}
\   \
\hfill \begin{minipage}{7cm}
\begin{center}
\includegraphics[width=7.5cm]{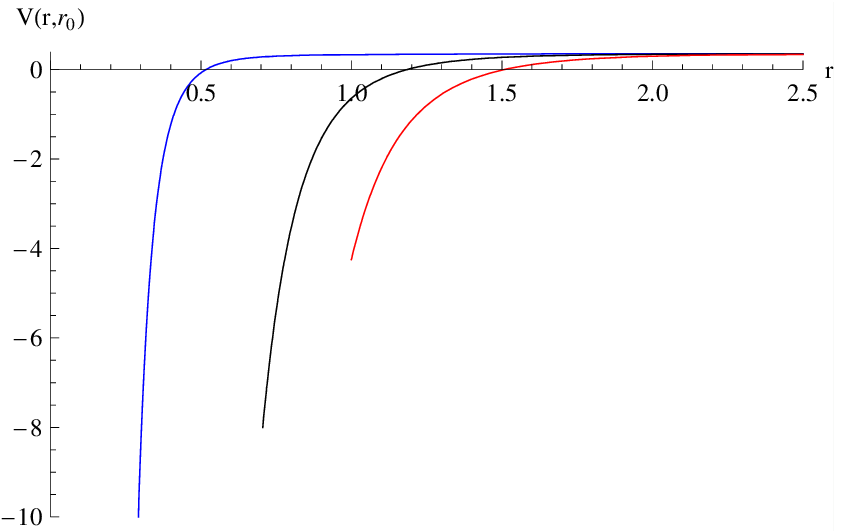}
\caption{Schrodinger potential for in-plane fluctuations depending on $t,r$-coordinates
as a function of $r_0$ in the Maldacena-N\'u\~nez t Hooft loop case. Blue, black and red lines correspond to
$r_0=0.2,\,0.7, 1$.}
\label{potschrMNthooft}
\end{center}
\end{minipage}
\end{figure}

In figures \ref{lvsRminMNthooft}  we plot the behavior
of the length function (\ref{generalL}) as a function of $r_0$. The length function  is an increasing function of $r_0$ and from the previous discussions
we therefore expect the embedding to be unstable.
The instability of the embedding can be easily seen in the Maldacena-N\'u\~nez case since a fluctuation along the
$x_1$-direction depending only on $t,r$ is decoupled from the angular ones (consistent fluctuation). The $\delta x_1$
equation of motion results in (\ref{SL})  with $f(r)=g(r)= h(r)=r\tanh r\, e^{\phi(r)}$.
The asymptotics of the fluctuation is the same as in the Wilson loop case, nevertheless the behavior
drastically changes near the origin since $f(r)$ goes to zero. As seen in figure \ref{wvsrminMNthooft}
negative eigenvalues exist for all $r_0$ values. For completeness we plot in figure \ref{potschrMNthooft}
the Schrodinger potential associated with the in-plane fluctuation equation of motion.

In figure \ref{EvsLMNthooft} we plot the energy as a function of the endpoints separation length $L$.
The energy of the configuration is positive for all $L$, this fact and the instability of the configuration
suggests that the stable configuration for given boundary conditions is the one corresponding to two ``straight lines''.
Contrary to the Wilson loop case the ``straight lines'' (used as reference state for regularizing the energy)
can end at the origin since they correspond to wrapped $D3$ on the topological $S^2$ of (\ref{metric})
which smoothly collapses at the origin.

\subsection{Klebanov-Strassler}

\begin{figure}[ht]
\begin{minipage}{7cm}
\begin{center}
\includegraphics[width=7.5cm]{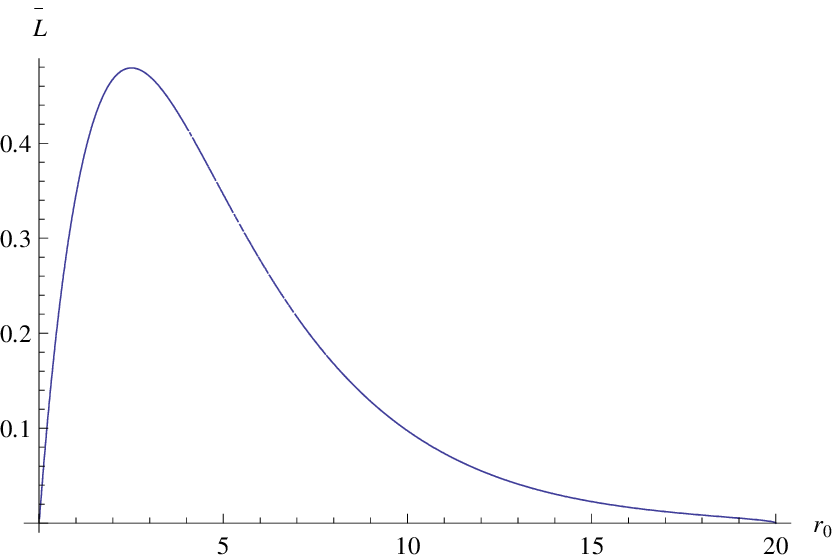}
\caption{Length function $L(r_0)$ for the effective string as a function of $r_0$ in KS background.}
\label{lvsRminKSthooft}
\end{center}
\end{minipage}
\   \
\hfill \begin{minipage}{7cm}
\begin{center}
\includegraphics[width=7.5cm]{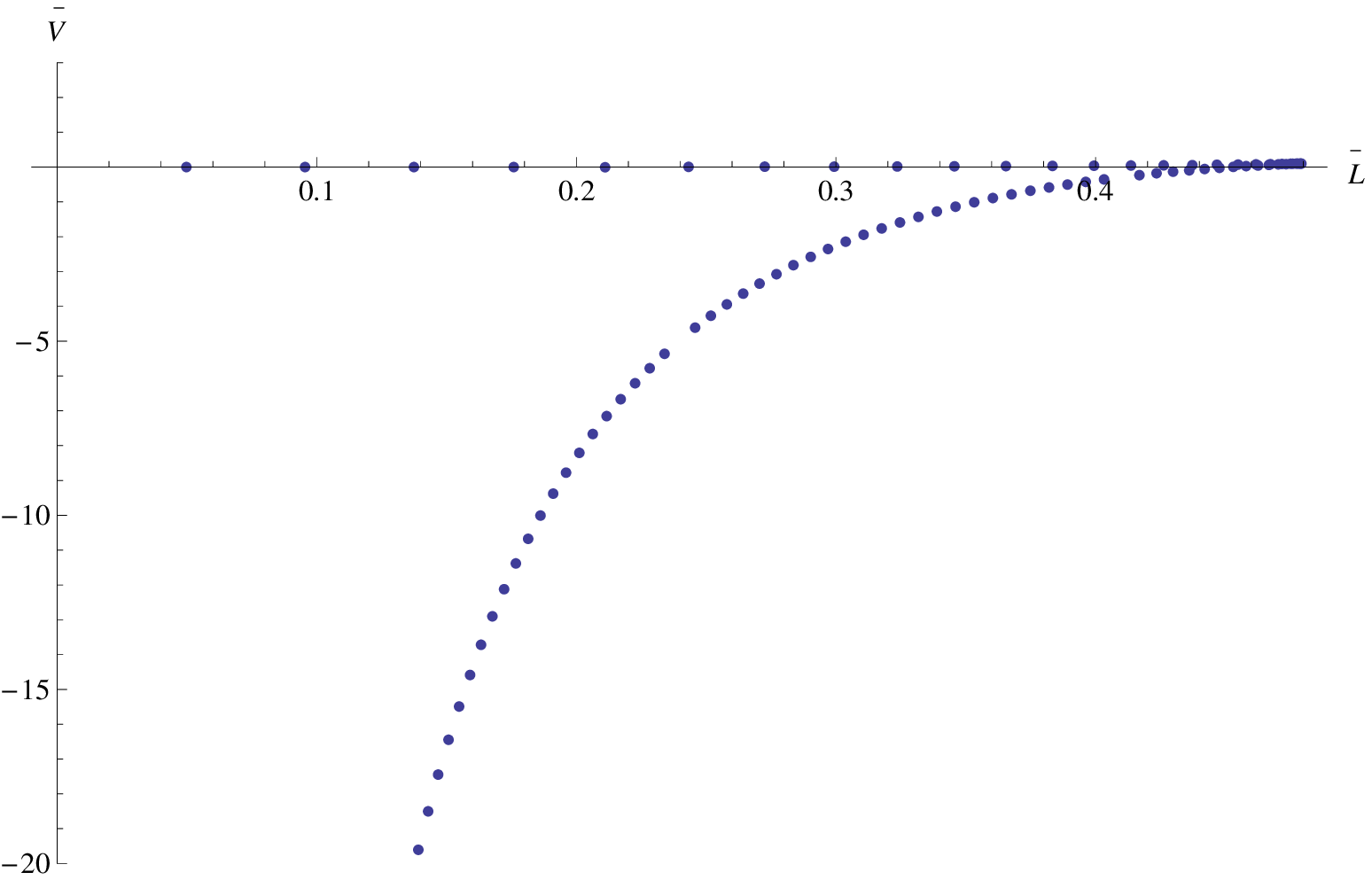}
\caption{ The energy of the monopole-antimonopole pair as a function of the separation length.}
\label{EvsrminthooftKS}
\end{center}
\end{minipage}
\end{figure}

This case again involves wrapping a $D3$ over a topological $S^2$ inside (\ref{ks})  (for its parametrization see the
appendix A of \cite{ho}). There are important differences with respect to the MN case, in the present case
the $H_3$ supporting the KS geometry contributes to the string action (\ref{NG}), and moreover it
leads to the entanglement of the angular and the in-plane fluctuations. The different UV behavior wrt the MN
is the reason for the length function $L(r_0)$ having a priori stable regimes
(see figure \ref{lvsRminKSthooft}).
The behavior of $\bar V(\bar L)$ shows the potential is screened for large $L$ and this agrees with the linear
confinement potential for the Wilson loop case (see \cite{groo} for a related example).
The analysis of the present case is analogous to the one in sect. \ref{adssch}, summarizing
when the configuration energy becomes positive, the two ``straight lines'' solution becomes favored.
As for the MN case this becomes possible without a horizon in this case due to the $D3$ being wrapped
over a smoothly collapsing $S^2$ at the origin. We did not attempt the analysis of the coupled fluctuations
equations of motions to check for instabilities on the left branch of fig. \ref{lvsRminKSthooft}.

\section{Conclusions}
\label{concl}

In this work we have analyzed the string proposal for computing rectangular Wilson loops via string
embeddings in gravity backgrounds and we have studied their stability under linear perturbations.

\vspace{1.5mm}

The string prescription involves solving for a minimal open string
worldsheet whose endpoints lie on the loop to be computed located at
a fixed value of the holographic radial direction. When the
endpoints are moved to infinity a divergent area results and a
regularization is mandatory in order to get a meaningful answer. In
section \ref{wil} we have reviewed this prescription and showed how
a finite value is obtained. We have chosen to regularize the action
by the standard procedure originally  proposed in
\cite{maldawilson}. This is interpreted as saying that the Nambu-Goto
worldsheet area computation includes the interaction energy plus the
self energy (mass) of the external quarks. Within this
interpretation we reproduced the well known results for $AdS$ and
thermal $AdS$. The regularization was in fact responsable for
turning the original positive area into a negative attractive
potential energy. When turning to smooth backgrounds ($AdS$ in
global coordinates, MN and KS) a puzzle arises  since the straight strings running along the radial
direction used in the substraction prescription must end somewhere in the bulk.
We concluded that the
correct interpretation for the substraction procedure is that we are
comparing the string `Wilson loop' worldsheet with respect to a
reference state consisting in a straight string worldsheet whose
endpoints lie at antipodes of a compact direction (pictorically
represented in fig. \ref{substraction}). It then follows that the
reference state in general satisfies different boundary conditions
than the worldsheet used in computing the expectation value for the
rectangular Wilson loop. This last observation is welcomed for the
MN and KS cases where the  linear confining relation occurs for
worlsheets having positive regularized energies (see figs.
\ref{EwilsonMN} and \ref{EwilsonKS}): if the reference state
satisfied the same boundary conditions as the `Wilson loop'
worldsheet, the observed linear behavior should not be considered
since the reference state ($E_{q\bar q}=0$) would have been the lowest energy
one (cf. last paragraph of sect. \ref{adssch}), but from the
previous analysis we see that this is not the case. We would like to
recall an observation in \cite{groo} stating that the relation between
Wilson loops and strings in gravity duals (at the semiclassical
level)
\be
\langle W\rangle\simeq e^{-A}
\ee
is schematic since the addition of boundary
terms to the Nambu-Goto action does not change the
minimal area character of the solutions but turns the value of the
classical action into something different than the area. In \cite{groo}
this arbitrariness was used to make a Legendre transform of the Nambu-Goto
action showing that the resulting quantity, for the case of loops
in $AdS$, is free from the linear divergences arising from the behavior of the worldsheet
near the boundary of $AdS$.

\vspace{1.5mm}

We also discussed the concavity conditions (\ref{convexity}) that must be satisfied
by any potential pretending to describe the
interaction between physical quarks. Generic gravity duals have positive and increasing $f(r)$ functions, so the
concavity conditions are not satisfied when the
length function is an increasing function of the minimal radial position reached
by the string $r_0$. In section \ref{bkg} we performed the analysis of the
length and potential  functions $L(r_0)$ and $V_{\sf string}(L)$ for different
backgrounds and showed that some of them lead to embedding solutions where the concavity condition fails.

Based on previous work \cite{pufu}-\cite{avramis} we studied linear fluctuations around
the embedding to test the stability of the classical embedding. We concluded that whenever the solution leads to an
unphysical potential, not satisfying the conditions (\ref{convexity}) there exist unstable
modes under linear fluctuations. In the course of the analysis we discussed the different
gauge fixings that can be imposed and its relation with the diffeomorphism of the Nambu-Goto action.
Three natural gauge fixings where discussed and we chose to work in the $r$-gauge since it lead to
simpler closed
expressions for the fluctuations equations of motion (see eqn. (\ref{lagr})). The
$r$-gauge leads to singular behavior
in the fluctuation at the tip of the embedding, but reviewing \cite{avramis} we showed that nevertheless
they are physical once an appropriate gauge transformation is performed.

In section \ref{stab} we perform the stability analysis for the solution reviewed in section \ref{bkg}.
We showed by a numerical analysis that the $AdS_5\times S^5$, Maldacena-N\'u\~nez and Klebanov-Strassler are stable. On other
hand for  thermal $AdS$ and the generalized Maldacena-N\'u\~nez backgrounds of sect. \ref{gmnsol}
we found unstable modes in agreement with the behavior of the $L(r_0)$ relation. This last case is
rather pathological since the loop cannot be placed at infinity and moreover we found that a minimum
separation exists beyond which no smooth solution connecting the string endpoints exists.
In section \ref{scho} transforming
the Sturm-Liouville fluctuation equations of motion into a Schrodinger like equation we reanalyzed
the problem, finding complete agreement with the results obtained in section \ref{stab}.
We conclude that the regions where we find unstable modes coincide with the regions where
the concavity condition fails.

\vspace{1.5mm}

In the last section we performed the previous analysis for the case of monopole-antimonopole interaction
in the non-conformal gravity duals of Maldacena-N\'u\~nez and Klebanov-Strassler.
We discussed the 't Hooft loop string prescription given by wrapping a $D3$ on the topological $S^2$
present in the geometries. The MN case was shown to be unstable for all $r_0$ values.  A fluctuation
analysis was feasible since a decoupled equation for the in-plane fluctuation could be found
were an unstable mode was shown to exist by a numerical analysis. The KS presented a behavior similar
to thermal $AdS$ with presumably stable and unstable regions, but a the fluctuation analysis lead to
coupled fluctuations equations of motion which we did not analyze.

We conclude that the analysis of Wilson/'t Hooft loops in given gravity background by looking at the
value of the $f^2$ at the origin should be supplemented with an analysis of the $L(r_0)$ relation.

\section*{Acknowledgments}

We thank D. Arean,   N. Grandi, A. Lugo, J. Maldacena,  M. Schvellinger,
M. Sturla for helpful discussions
and correspondence. We are specially grateful with C. Nu\~nez for a careful reading of
the manuscript and sharing with us a draft of \cite{piai}.
This work was partially supported by PIP6160-CONICET and by the ANPCyT PICT-2007-00849.

\appendix

\section{Sturm-Liouville to Schr\"odinger}
\label{sl2sc}

Equations (\ref{SL})-(\ref{spp}) are of the Sturm-Liouville type
\be
\Big[-\frac{d}{dr}\left(P(r,r_0)\frac{d}{dr}\right)+U(r,r_0)\Big]
\Phi(r)=\omega^2Q(r,r_0)\Phi(r),~~~ r_0 \leq r<\infty
\label{sliouv2}
\ee
the functions $P(r,r_0)$ and $Q(r,r_0)$  can be read off
from (\ref{SL})-(\ref{spp}), $U(r,r_0)=0$ in both cases. The change of variables
\be
y=\int_{r_0}^r \sqrt{\frac{Q }{P }}\;dr,~~~~
\Phi(r)=\left(PQ\right)^{-\frac14}\Psi(y)
\label{chv}
\ee
transforms (\ref{sliouv2}) to a
Schr\"odinger like equation
\be
\left[-\frac{d^2}{dy^2}+V \right]\Psi =\omega^2\Psi ,~~~ 0 \leq y\leq y_0\,.
\label{sch}
\ee
Here $y_0=\int_{r_0}^\infty dr\sqrt{\frac{Q}{P}}$ which may be finite or infinite depending on the nature of $Q,P$ and
one can check that (\ref{chv}) is integrable at the lower limit giving $y\sim\sqrt{r-r_0}$.
The potential $V$
is given by
\bea
V&=&\frac U Q+\left[(P Q )^{-\frac14}\frac{d ^2}{dy^2}\right](P Q )^{\frac14}\nn\\
&=&\frac U Q+\left[\frac{P ^{\frac14}}{Q ^{\frac34}}\frac{d}{dr}\left(\sqrt{\frac{P}{Q}}\frac{d}{dr}\right)\right]
(PQ)^{\frac14}
\label{schrodinger}
\eea
The points  $r=r_0$ and $r=\infty$  map to $y=0$ and $y=y_0$ respectively. The boundary conditions
to be imposed on the solutions of (\ref{sch}) are \cite{avramis}:
\begin{itemize}
\item Infinity: string endpoints fixed\footnote{See however a loophole in the Maldacena-N\'u\~nez  context
(sect. \ref{schMN}) when
imposing (\ref{bcsch}).}
\be
\delta x|_{r=\infty}=0 \Rightarrow\Psi|_{y=y_0}=0
\label{bcsch}
\ee

\item Tip $r=r_0$: for both in-plane $\delta x_1$ and transverse fluctuations
$\delta x_m$  one obtains from (\ref{xperp}),(\ref{bc}),(\ref{bc2})
\be
\begin{array}{c}
  ~\mathrm{Even~solutions}:\left.\frac{d\Psi}{dy}\right|_{y=0}=0 \\
  \mathrm{Odd~solutions}: ~~\left. \Psi \right|_{y=0}=0\,.
\end{array}\label{bcsschrod}
\ee
 \end{itemize}

\section{Exact spectrum for transverse fluctuations in $AdS_5\times S^5$}
\label{kmt}

We review here the solution of \cite{kmt} for the exact spectrum of the longitudinal
fluctuations in the $AdS_5\times S^5$ background and compare it with our numerical results
using the shooting technique described at the end of section \ref{stability}.

The $AdS$ metric is written in Poincare coordinates
\be
ds^2=\frac{R^2}{z^2}(-dt^2+dx_{i}dx_{i}+dz^2)+R^2d\Omega_5^2\, .
\ee
A $x$-gauge fixed ansatz $t=\tau, x=x_{\sf cl}, z=z_{\sf cl}(x)$ leads to
\be
\left(\frac {dz_{\sf cl}}{dx} \right)^2=\frac{z_0^4-(z_{\sf cl})^4}{(z_{\sf cl})^4}\,.
\label{zcl}
\ee
The solution to (\ref{zcl}) with the string endpoints separated by a distance $L$ is (\ref{xads})-(\ref{lads}).
\be
x_{\sf cl}(z)=\pm z_0\left[\frac{(2\pi)^{\frac32}}{2\Gamma[\frac14]^2}
-\frac14\mathsf{B}\left( \frac{z^4}{z_0^4} ;\frac34,\frac12\right)\right]
\label{adssol}
\ee
where $z_0=z_{\sf cl}(0)=({\Gamma[\frac14]^2}/{(2\pi)^{\frac32}})L$ is the maximal radial distance reached by the string (tip
of the string).

Fluctuations around the solution (\ref{adssol}) in the transverse $x_m$ ($m=2,3$) coordinates decouple, writing
$X^\mu=(t,x_{\sf cl}(\sigma),\delta x_m(t,\sigma),z_{\sf cl}(\sigma))$ the
equations  to linear order are \cite{kmt},\cite{cg}
\bea
x\mathrm{ -gauge}:&&\left[\partial_t^2-\frac{z_{\sf cl}^4(x)}{z_0^4}\,\partial_x^2\right]\delta x_m(t,x)=0\label{xg}\\
r\mathrm{ -gauge}:&&\left[\partial_t^2-(1-\frac {z^4}{z_0^4})\, \partial_z^2+\frac2z\,\partial_z\right]\delta x_m(t,z)=0~~~~~~~m=2,3\,.
\label{cg}
\eea
As mentioned in section \ref{wilson}, note that the $x$-gauge equation of motion (\ref{xg}) depends explicitly on the
classical solution $z_{\sf cl}(x)$.  The equations are related by the change of variables given in (\ref{adssol}). Writing
$\delta x_m=e^{-iwt} f(z)$  in (\ref{cg}) and calling $\tilde z=z/z_0$ one obtains \cite{cg}
\be
\left[(1-\tilde z^4)\partial_{\tilde z}^2 -\frac2{\tilde z}+\xi^2\right]f(\tilde z)=0,~~~~~0\le \tilde z\le 1\,,
\label{xads2}
\ee
where $\xi=z_0\omega$. The change of variables \cite{btt}
\bea
f(\tilde z)&=&\sqrt{1+\xi^2\tilde z^2}F(q)\nn\\
q(\tilde z)&=&\pm 2\int_{\tilde z}^1\frac {t^2}{(1+(\xi
t)^2)\sqrt{1-t^4}}\,dt
\eea
transforms equation (\ref{xads2}) into a
simple harmonic oscillator \be \frac
{d^2F}{dq}+\frac14\,\xi^2(\xi^4-1)F=0,~~~~q\in[-q_*,q_*]
\label{osci} \ee where $q_*=q(0)$. The boundary conditions at
infinity $\delta x_m(t,0)=0$ have been mapped to $F(q_*)=0$, and
quantize the frequencies in (\ref{osci}) leading to \be
\omega_nz_0\sqrt{\omega_n^4z_0^4-1}\int_0^1\frac{t^2dt}{(1+w_n^2z_0^2)\sqrt{1-t^4}}=\frac{n\pi}{2},~~~~n=1,2,...\label{exact}
\ee The following table shows the comparison between the exact
eigenvalues (\ref{exact}) and our numerical calculation of eigenvalues of (\ref{xads2}) with $z_0=1$.
\bc
\begin{tabular}{|r||c|l|}
 \hline
   & {Exact} & {Numeric} \\
     \hline
  $\omega_1$ & 2.203 & ~2.226\\
  $\omega_2$ & 3.467 & ~3.492\\
  $\omega_3$ & 4.697 & ~4.735\\
  $\omega_4$ & 5.914 & ~5.959\\
  $\omega_5$ & 7.125 & ~7.181\\
  $\omega_6$ & 8.332 & ~8.396\\
  $\omega_7$ & 9.537 & ~9.612\\
  $\omega_8$ & 10.741 & ~10.823\\  \hline
\end{tabular}
\ec
The odd (even) eigenvalues where obtained solving (\ref{xads2}) with the even (odd) boundary conditions
discussed after (\ref{xperp}).

\end{document}